\documentclass[prb,superscriptaddress,amsmath,amssymb,footinbib,twocolumn]{revtex4-2}
\pdfoutput=1
\usepackage{physics}
\usepackage{lipsum}
\usepackage{soul}
\usepackage{natbib}
\usepackage[english]{babel}
\makeatother
\usepackage{layouts} % measure linewidth
\usepackage{graphicx}
\usepackage[usenames,dvipsnames]{color}
\usepackage{stackengine}
\usepackage[T1]{fontenc}
\usepackage[colorlinks=true,allcolors=blue]{hyperref}

\begin{document}

\title{One-dimensional non-interacting topological insulators with chiral symmetry}
\author{Polina Matveeva}
\affiliation{Department of Physics, Bar-Ilan University, Ramat Gan, 52900, Israel}
\author{Tyler Hewitt}
\affiliation{School of Physical Sciences, University of Kent, Canterbury CT2 7NH, United Kingdom}
\author{Donghao Liu}
\affiliation{Department of Physics, Bar-Ilan University, Ramat Gan, 52900, Israel}
\author{Kethan Reddy}
\affiliation{School of Physical Sciences, University of Kent, Canterbury CT2 7NH, United Kingdom}
\author{Dmitri Gutman}
\affiliation{Department of Physics, Bar-Ilan University, Ramat Gan, 52900, Israel}
\author{Sam T.~Carr}
\affiliation{School of Physical Sciences, University of Kent, Canterbury CT2 7NH, United Kingdom}
\date{\today}

\begin{abstract}
We construct microscopical models of one-dimensional non-interacting topological insulators in all of the chiral universality classes.
Specifically,  we start with a deformation of the Su-Schrieffer-Heeger (SSH) model that breaks time-reversal symmetry, which is in the AIII class.  We then couple this model to its time-reversal counterpart in order to build models in the classes BDI, CII, DIII and CI.   We find that the $\mathbb{Z}$ topological index (the winding number) in individual chains is defined only up to a sign.  This comes from noticing that changing the sign of the chiral symmetry operator changes the sign of the winding number.
The freedom to choose the sign of the chiral symmetry operator on each chain independently allows us to construct two distinct possible chiral symmetry operators when the chains are weakly coupled -- in one case, the total winding number is given by the sum of the winding number of individual chains while in the second case, the difference is taken.
We find that the chiral models that belong to $\mathbb{Z}$ classes, AIII, BDI and CII are topologically equivalent,  so they can be adiabatically deformed into one another so long as the chiral symmetry is preserved.  We study the properties of the edge states in the constructed models and prove that topologically protected edge states must all be localised on the same sublattice (on any given edge).  We also discuss the role of particle-hole symmetry on the protection of edge states and explain how it manages to protect edge states in $\mathbb{Z}_2$ classes, where the integer invariant vanishes and chiral symmetry alone does not protect the edge states anymore. 
We discuss applications of our results to the case of an arbitrary number of coupled chains, construct possible chiral symmetry operators for the multiple chain case, and briefly discuss the generalisation to any odd number of dimensions. 
\end{abstract}

\maketitle
\section*{Introduction} 

Topology plays a central role in diverse parts of physics,
ranging from superfluid He to elementary particles \cite{Thouless, Haldane2017, Volovik}.  In the context of electronic solid-state
physics, the topological numbers first arose in the integer quantum Hall effect. The perpendicular
magnetic field breaks time-reversal symmetry and induces Landau quantization that opens
a gap in the bulk spectrum of a two-dimensional metal. The closing of the gap at the boundary
is imposed by the non-trivial first Chern number \cite{Thouless1982} and leads to the protected chiral modes on
the edge of the sample. The topological nature of these modes renders them robust against
disorder and enforces conductance quantization. 

It was more than 20 years later before it was realised that the Chern number was only one of many possible topological invariants that could be used to classify phases in gapped non-interacting electronic systems.  The key was in identifying the role of symmetry -- the quantum Hall state is indeed the only state (in fewer than four spatial dimensions) that truly cannot be smoothly deformed to a trivial insulator.  However, if one imposes a symmetry such as time-reversal, then it is possible to have different topologically distinct phases that can not be deformed to the trivial insulator without breaking the time reversal symmetry or closing the gap.  Such a state was first proposed by Kane and Mele in 2005 for a two-dimensional system without a magnetic field but with strong spin-orbit interactions \cite{Kane2005}. More realistic models of materials \cite{Bernevig2006, Bernevig22006} along with experimental verification \cite{Koenig2006} followed soon afterwards.  In these systems, the edge modes are helical rather than chiral, and only protected against disorder scattering so long as time reversal symmetry remains unbroken.  It was also quickly realised that, unlike the quantum hall state, this new time reversal symmetric topological insulator has an analogue in three dimensions, and much theoretical and experimental work followed \cite{Fu2007,Qi2008,Hsieh2008,Koenig2008,Roth2009,Hsieh2009,Hsieh20092,Wang2010,Hasan2010,Qi2011}.

Alongside this work on specific materials and models, a classification scheme emerged \cite{Schnyder2008,Kitaev2009} on what topological indices are possible in given systems.  It was discovered that this depended only on the number of spatial dimensions $d$, along with which symmetries were present or not in the system -- time reversal symmetry, particle hole symmetry, or the combination of the two known as chiral symmetry.  The topological indices were either $\mathbf{Z}$, as in the Chern number in the quantum Hall effect, or $\mathbf{Z}_2$ as in the newly discovered time-reversal symmetric topological insulators in two and three dimensions.

Within this classification scheme, one-dimensional (1D) models  are special  because they can be realized  experimentally, and at the same time are simple enough to allow an intuitive picture to emerge.
The best-known model  of a one-dimensional  topological insulator is the Su-Schrieffer-Heeger (SSH) model \cite{Su1979,RyuHatsugai2006} of polyacetylene. 
Similar models  were recently realized in experiments with ultracold atoms \cite{Bloch2013,Leder2016}.
In addition, a  toy model of a one-dimensional topological superconductor was proposed by Kitaev \cite{Kitaev_2001}. It describes the simplest one-dimensional p-wave superconductor and turns out to have the same Hamiltonian as the SSH model with the fermions replaced by Majorana fermions.
 This model hosts zero-energy Majorana edge modes that are discussed in the context of topologically protected quantum computation \cite{Nayak2008,beenakker2019search}.
 
 Both the SSH model and the Kitaev model fall in the universality class BDI which has a $\mathbf{Z}$ topological index in one dimension.  This class has chiral symmetry, along with time-reversal symmetry and particle-hole symmetry.   In this paper we investigate how one may use the SSH model as a building block to construct toy models in other universality classes.
 
This follows a very similar recipe that was used by Kane and Mele\cite{Kane2005} to construct the first model in 2D of a time-reversal symmetric topological insulator.  One begins with two copies of an earlier model by Haldane \cite{Haldane1988}.  One of these models has Chern number of $+1$ while the other is the time-reversal counterpart of the first and has Chern number of $-1$.  They are coupled together in a way that then respects time-reversal symmetry.  This coupled model has a Chern number of $0$, however in this case one can form a different $\mathbf{Z}_2$ index characterising the topological nature of the state.

In our case, our starting point is a deformation of the SSH model that breaks time-reversal and particle-hole symmetries, leaving only chiral symmetry (this is the AIII universality class).  We then couple such a model to its time-reversed counterpart to create a new model with time reversal symmetry restored.  There is a major difference however between following this procedure in two dimensions and one: in one dimension, we deal with the \textit{winding number}, not the \textit{Chern number}.  While the Chern number is odd under time-reversal, the winding number is not.  This means that the two subcomponents of the model before coupling both have a winding number $\nu=1$.  However rather surprisingly,  once the two models are weakly coupled (without closing the gap and retaining chiral symmetry),  the winding number is not necessarily $2$ -- it is also possible for it to be $0$ depending on exactly how the two models are coupled.

The reason for this unexpected result is rather simple -- the winding number is odd under a relabelling of the atoms within the unit cell.  We will show in the next section that this relabelling can be equivalently thought of as changing the sign of the chiral symmetry operator.  As this labelling is entirely arbitrary and has no physical consequences, the winding number is only defined \textit{up to a sign} -- in other words, there is no difference between winding number $+1$ and $-1$.   This means that if one takes two models with winding numbers $\nu_1$ and $\nu_2$,  there are two distinct classes of allowed coupling terms respecting chiral symmetry.  One of these will lead to a resulting total winding number $\nu_\text{tot}=\nu_1 + \nu_2$, while the other will give $\nu_\text{tot}=\nu_1 - \nu_2$.  
 
In this paper, we derive and explain this sign ambiguity in the winding number.  We then use the procedure outlined above to construct simple 
microscopic models of non-interacting topological insulators 
and superconductors in all of the classes with chiral symmetry in one-dimension.
Though the topological properties of non-interacting gapped models
 have been fully classified,\cite{Schnyder2008, Kitaev2009} we find that certain simplifications occur.
 In particular, we show that systems that belong to the different chiral symmetry classes with a $\mathbf{Z}$ classification are in fact  topologically equivalent -- for example models in the BDI class have time-reversal and particle-hole symmetry in addition to the chiral symmetry,  however,  any deformation that respects the chiral symmetry (but may break time-reversal and particle-hole) will not gap the zero-energy edge states.  

While we derive such results for specific models, we also show they are true in general in any odd dimension.
It is worth mentioning, that similar results were obtained in the context of disordered quasi-one-dimensional wires \cite{AltlandKamenev2015, AltlandKamenev2014}.
It was shown that for all chiral classes the RG flows to the same fixed point, with an identical number of protected 
 edge states and the Anderson insulating bulk. The particular type of the chiral class plays no role at the infrared fixed point.

We discuss the general properties of the edge states in chiral symmetric systems with integer classification.  We show in this case that the edge states are localised on only one of the sublattices (on any given edge), and furthermore if there are multiple protected edge states, they must all be localised on the same sublattice. 
We also discuss the role of particle-hole symmetry in the protection of the edge states,  and demonstrate that due to anti-unitarity, the particle-hole symmetry protects either none or only one from an odd number of edge states,  depending on whether it squares to -1 or +1 correspondingly.  We discuss how it is related to the $\mathbb{Z}_2$ classification in the D and DIII classes in 1D. 

The paper is organized as follows:  In Section \ref{Section1} we briefly review the topological classification of one-dimensional systems and review an analytical expression for the winding number that allows us to compute the index without diagonalizing the Hamiltonian. 
We also discuss the sign ambiguity of the winding number and topological equivalence of chiral symmetric models belonging to different classes. 
In Section \ref{section2} we illustrate our ideas by explicitly constructing minimal real-space models of chiral topological insulators by coupling two SSH models.  We compute their topological indices in case of coupling weak enough that it doesn't close the gap.  And we also discuss the symmetry properties of  edge states in those models.  In Section \ref{section3} we focus on the general properties of the edge states in chiral symmetric systems. We also discuss the edge states in the presence of particle-hole symmetry. 
In Section \ref{section4} we generalize our results for the multiple coupled chains and evaluate the winding number in the weak coupling limit.

\section{Winding number of one-dimensional gapped systems.}\label{Section1}
Here we briefly discuss the general aspects of topological classification of non-interacting topological phases important for the rest of the paper. All non-interacting topological systems fall into 10 topological classes, depending on three symmetries. These are time-reversal symmetry $T$, particle-hole symmetry $P$ and chiral symmetry $C$.\cite{Zirnbauer1996,Altland1997} In the single-particle Hilbert space, symmetries $T$ and $P$ are anti-unitary,  and thus they can be represented as $T=U_T\mathcal{K}$ and $P=U_P\mathcal{K}$, where $U_T$ and $U_P$ are unitary matrices and $\mathcal{K}$ is complex conjugation.  

By definition, the system with the Hamiltonian $H$ possesses these symmetries if the following holds: 
\begin{align}
\label{symmetries}
\text{Time-reversal:} \hspace{0.3cm} U^{-1}_T H^*(-\mathbf{k}) U _T=H (\mathbf{k}) \nonumber \\ 
\text{Particle-hole:} \hspace{0.3cm} U^{-1}_P H^*(-\mathbf{k}) U _P=-H (\mathbf{k}) \\ 
\text{Chiral:} \hspace{0.3cm} U^{-1}_C H(\mathbf{k}) U _C=-H (\mathbf{k}). \nonumber
\end{align}
Note, that as time-reversal and particle-hole symmetries are anti-unitary they square either to $+1$ or $-1$. If the system possesses both $P$ and $T$ symmetries it is also chiral symmetric $C=P\cdot T$. Thus, depending on the type of time-reversal and particle-hole symmetries, there are four chiral classes: BDI, CI, CII, and DIII, see  Table \ref{chiral_classes1D}. However, as the presence of chiral symmetry alone does not guarantee that a system is particle-hole and time-reversal symmetric, there is also a class AIII that possesses only chiral symmetry.   Like time-reversal and particle-hole symmetries, applying the chiral symmetry operator twice does not change the state so is proportional to the identity.  However,  as the chiral symmetry is unitary as opposed to anti-unitary,  one can always remove the phase here by a gauge choice -- so $U_C^2=1$ by definition -- see Ref.~\onlinecite{Ludwig_2015} for more details.

While fixing $U_C^2=1$ is almost enough to uniquely define the chiral symmetry operator $U_C$, there is still a sign ambiguity as $U_C \rightarrow -U_C$ clearly also satisfies this condition.  The essence of this paper is exploring the consequences of this ambiguity.

\begin{table}
\begin{tabular}{|c|c|c|c||c|}
\hline
 Class & $T^2$ & $P^2$ & $C$& Topological Index in 1D \\
 \hline
 AIII & 0 & 0 & 1&$\mathbb{Z}$ \\ \hline
BDI & 1 & 1 &1 &$\mathbb{Z}$ \\
CII & -1 & -1 & 1 & 2$\mathbb{Z}$ \\
DIII & -1 &1 & 1 &$\mathbb{Z}_2$ \\
CI & 1 & -1 & 1&0 \\
\hline
\end{tabular}
\label{chiral_classes1D}
\end{table}
 
The chiral symmetry is equivalent to a sub-lattice symmetry.  To see this, define two operators
\begin{equation}
\label{sublatticeprojection}
P_{A(B)} = \frac{1}{2} \left( 1 \pm U_C \right).
\end{equation}
These satisfy $P_A+ P_B = 1$ and $P_A P_B = 0$, therefore they may be considered as projection operators onto $A$ and $B$ sub-lattices under a bi-partite division of the lattice (in some basis).  If the chiral symmetry operator is not diagonal, this natural basis for the sub-lattices may not be physically transparent, however the bipartite sub-lattice division is always present in models with chiral symmetry.  We also note that under a change of sign of the chiral symmetry operator $U_C \rightarrow -U_C$, the labelling of the sub-lattices swap $P_A \leftrightarrow P_B$ which follows directly from Eq.~\eqref{sublatticeprojection}.  Depending on the context, it is sometimes more useful to think of a change of sign of the chiral symmetry operator and other times better to think of a re-labelling of the sublattices.  Having shown these are equivalent, we will use these two descriptions interchangeably throughout this work.

The anti-commutation of $U_C$ with the Hamiltonian Eq.~\eqref{symmetries} in models with chiral symmetry implies
\begin{equation}
P_A H P_A = P_B H P_B = 0.
\end{equation}
Hence in this basis, there is no direct coupling through the action of the Hamiltonian from $A$ sites to $A$ sites or $B$ sites to $B$ sites.  This means the Hamiltonian written in momentum space may be written in a block off-diagonal form:
\begin{align}
\label{chiral_ham}
\hat{H}=\sum_k c^{\dagger}_k \hat{h} (k) c_k,\\ 
\hat{h}(k)=\begin{pmatrix}
0 & \hat{\Delta}(k) \\
\hat{\Delta}^{\dagger}(k) & 0
\end{pmatrix},
\end{align}

The \textit{winding number} may then be defined via the formula: 
\begin{equation}
\label{phase_detq}
\nu=\frac{-i}{2\pi}\int\limits_{\text{BZ}} \partial_k i\phi(k) dk,
\end{equation}
where $\phi(k)$ is defined through the expression
\begin{equation}
\label{det-delta}
\det \Delta (k) = r(k) e^{i \phi(k) },
\end{equation}
i.e. $\phi(k)$ is the complex phase of $\det \Delta (k)$.
This expression is well known for the case of two-bands (see e.g. Ref. \onlinecite{Asboth_2016}) where the determinant is not required as $\Delta(k)$ is just a number. In Appendix \ref{AppendixA}, we show that by adding the determinant this expression is equivalent to that derived via $Q$ operators in Ref.~\onlinecite{Ryu2010} for a multi-band system.  We also note that instead of using $i\phi(k)$ in Eq.~\eqref{phase_detq}, one could replace this with $\ln \det \Delta(k)$ as the real parts will always cancel in the integral -- however we find it physically more intuitive to think of this integral as involving the phase only.

Thus, the value of topological index $\nu$ for a concrete model is determined by the number of poles of $f(k)=\partial_k \phi(k)$ as a function of a complex variable $z=e^{ik}$, that lie inside a unit circle -- hence the name `winding number' as it counts how many times $\det \Delta (k)$ winds around the origin in the complex plane. This formula allows for computing the topological index without diagonalizing Hamiltonian and thus may simplify analytical computations of topological invariants in concrete systems.

Clearly,  this index cannot change via adiabatic deformation without $\det \Delta (k)=0$ for some value of $k$ -- which would correspond to the gap closing.  Hence this index is a $\mathbf{Z}$ topological index protected by chiral symmetry.  The chiral symmetry is important, as the definition of $\nu$ requires the off-diagonal structure of the Hamiltonian in the sub-lattice basis, which in turn requires chiral symmetry.  Note however that no other symmetry plays a direct role in this definition -- as long as chiral symmetry is present, $\nu$ cannot be changed via any adiabatic deformation that doesn't close the gap.

This discussion has been specific to one dimension, however the arguments may be generalised to all \textit{odd} dimensions -- some details of this are described in Appendix \ref{AppendixB}.

\subsection*{Sign ambiguity}

By relabelling sublattice A as B and vice versa, the Hamiltonian \eqref{chiral_ham} becomes
\begin{align}
\label{chiral_ham2}
\hat{h}(k)=\begin{pmatrix}
0 & \hat{\Delta}^\dagger(k) \\
\hat{\Delta}(k) & 0
\end{pmatrix},
\end{align}
Under this transformation, the $\det \hat{\Delta}(k)$ in Eq.~\eqref{det-delta} becomes $\det \hat{\Delta}^\dagger(k)=(\det \hat{\Delta}(k))^*$.  This changes the complex phase $\phi(k) \rightarrow -\phi(k)$ and hence the winding number changes sign
\begin{equation}
\nu \rightarrow -\nu.
\end{equation}
As this change of labels has no physical consequences, we can conclude that \textit{the winding number $\nu$ is defined only up to a sign}.   In Appendix \ref{AppendixB}, we show that the same is true in all odd dimensions where the winding number can be defined.

Let us make two comments here:
\begin{enumerate}
\item Note that the sign ambiguity of the $\mathbb{Z}$ topological index exists only in odd dimensions. 
In even dimension, the equivalent index is the \textit{Chern number} rather than the \textit{winding number}, and the sign of the Chern number is a physically measurable quantity.
For example in two dimensions,  the Chern number is equal
to Hall conductance -- its sign is uniquely defined. 
Moreover, by coupling two Chern insulators with the opposite topological numbers 
one always gets a system that has a total Chern number zero. 
This is a standard way to construct time-reversal  $\mathbb{Z}_2$ TIs in two dimensions \cite{Kane2005}.
\item There is a different well-known ambiguity related to topological indices in one-dimension.  Rather than using winding number, one could focus on the Zak phase \cite{Zak1989} which is the Berry phase of a particle in a path in $k$-space through the Brioullin zone.  The Zak phase is related to a physical quantity: the polarisation per unit cell \cite{Resta, Smith, Vanderbilt_book}.  The Zak phase is however not gauge-invariant, and the polarisation depends on the choice of unit cell \cite{Fuchs2021}.   For instance, we may consider a one-dimensional periodic chain consisting of positively and negatively charged ions with the charges $ \pm e$. The unit cell of such a system can be chosen such as a positively charged ion labeled as $A$ and a negatively charged ion as $B$. Then the polarization of the unit cell is $\vec{P}=e (\vec{r}_{B} -\vec{r}_{A})$, where the vectors $\vec{r}_{A/B}$ characterize the position of ions along the chain. If we now shift the position of the unit cell by $a/2$, where $a$ -- the distance between the atoms, we will get another unit cell with exchanged positive and negative ions, thus the polarisation of a unit cell switches its sign $-\vec{P}$.   However this is very different from the ambiguity in the sign of the winding number -- which exists for a given unit cell.
\end{enumerate}

In our work, we construct models of topological insulators with chiral symmetry by coupling two spinless SSH chains. The fact that the winding number is defined up to a sign suggests that there are two ways of coupling the chains while preserving chiral symmetry. We discuss them in the next section.  

\section{Microscopic models of one-dimensional chiral topological insulators}\label{section2}
\subsection{Uncoupled one-dimensional chains}
Before we discuss how to construct models that represent chiral topological classes by coupling two SSH chains, we first focus on topological properties of an uncoupled system, described by the following Hamiltonian in the basis $c^{\text{T}}_n=\left\{c_{A ,1,n}, c_{B,1,n},c_{A,2,n},c_{B,2,n} \right\}$: 
\begin{align}
\label{block_H0}
\hat{H}_0=\begin{pmatrix}
\hat{h}_{\text{SSH}} &0 \\
0& \hat{h}^* _{\text{SSH}}
\end{pmatrix},
\end{align} 
where $\hat{h}_{\text{SSH}}$ is a Hamiltonian of an SSH model, given by: 
\begin{align}
\label{H_SSH}
\hat{h}_{\text{SSH}} = w \sum_{n=1}^N \hat{c}^{\dagger}_{A,n} \hat{c}_{B,n} + v \sum_{n=1}^{N-1} \hat{c}^{\dagger}_{B,n} \hat{c}_{A,n+1} +\text{h.c.},
\end{align}
where amplitudes $w, v \in \mathbb{C}$.  If $w$ and $v$ are real, $\hat{h}_\text{SSH}$ has time-reversal symmetry and is in the class BDI -- conventionally when people talk about the SSH model in this context, it is this case they mean.  If however we allow $w$ and or $v$ to be complex, time-reversal symmetry is broken and $\hat{H}_\text{SSH}$ belongs to the AIII universality class.\cite{Velasco2017}  There is a slight subtlety here that even with complex $v$ and $w$, one can apply a gauge transformation $c_n \rightarrow e^{i\alpha n} c_n$ to make the hopping amplitudes real again.  In other words, even with complex $v$ and $w$, the model has time reversal symmetry, so long as one defines the time-reversal operator correctly.  One can remove this subtlety by adding longer range hopping terms.  However once two such models are coupled as we will do, the time reversal symmetry may be truly broken as well, therefore this subtlety is unimportant for our purposes.

The winding number (modulo sign) of a single SSH chain is $\nu=1$, if $|w|<|v|$ and $\nu=0$ if $|w|>|v|$. 
As was discussed in the previous section, the sign of the winding number can be switched by A $\leftrightarrow$ B relabeling of the sublattices.  One way of viewing this is as a `choice' of chiral symmetry operators -- in the former case we have $C=S_z$ acting in the sublattice basis; while in the latter we have $C=-S_z$.  From Eq.~\eqref{sublatticeprojection}, we see that this simple change of sign switches the projections onto the $A$ and $B$ sublattices.  Obviously,  for a single chain, this change of sign makes no difference -- however when multiple chains are coupled, one has this individual choice on each chain, and the relative sign of them does matter.

To illustrate this, we start with the Hamiltonian of two uncoupled chains \eqref{block_H0}.  If we \textit{choose} the chiral symmetry operator to have the same sign on both chains, the combined chiral symmetry may be written $C_1=S_z\sigma_0$ where $\sigma_0$ acts in chain space.  Using this chiral symmetry operator, the total winding number is $\nu_1+\nu_2$, which may be $0$ or $2$ depending on the relative magnitude of $|w|$ and $|v|$.  However, if we \textit{choose} the chiral symmetry operator to have the opposite sign on each chain,  the combined chiral symmetry may be written $C_2=S_z\sigma_z$.  In this case, the total winding number is $\nu_1-\nu_2$ which is always $0$ for the form of model we have chosen.  While it may seem disconcerting that the total winding number can not be uniquely defined, this is because the model is block diagonal -- i.e. has an additional unitary symmetry leading to individual conservation of charge in each chain.  When this symmetry is removed by coupling the chains, the winding number is once again uniquely defined (up to a sign).

\begin{figure}
\includegraphics[scale=0.23]{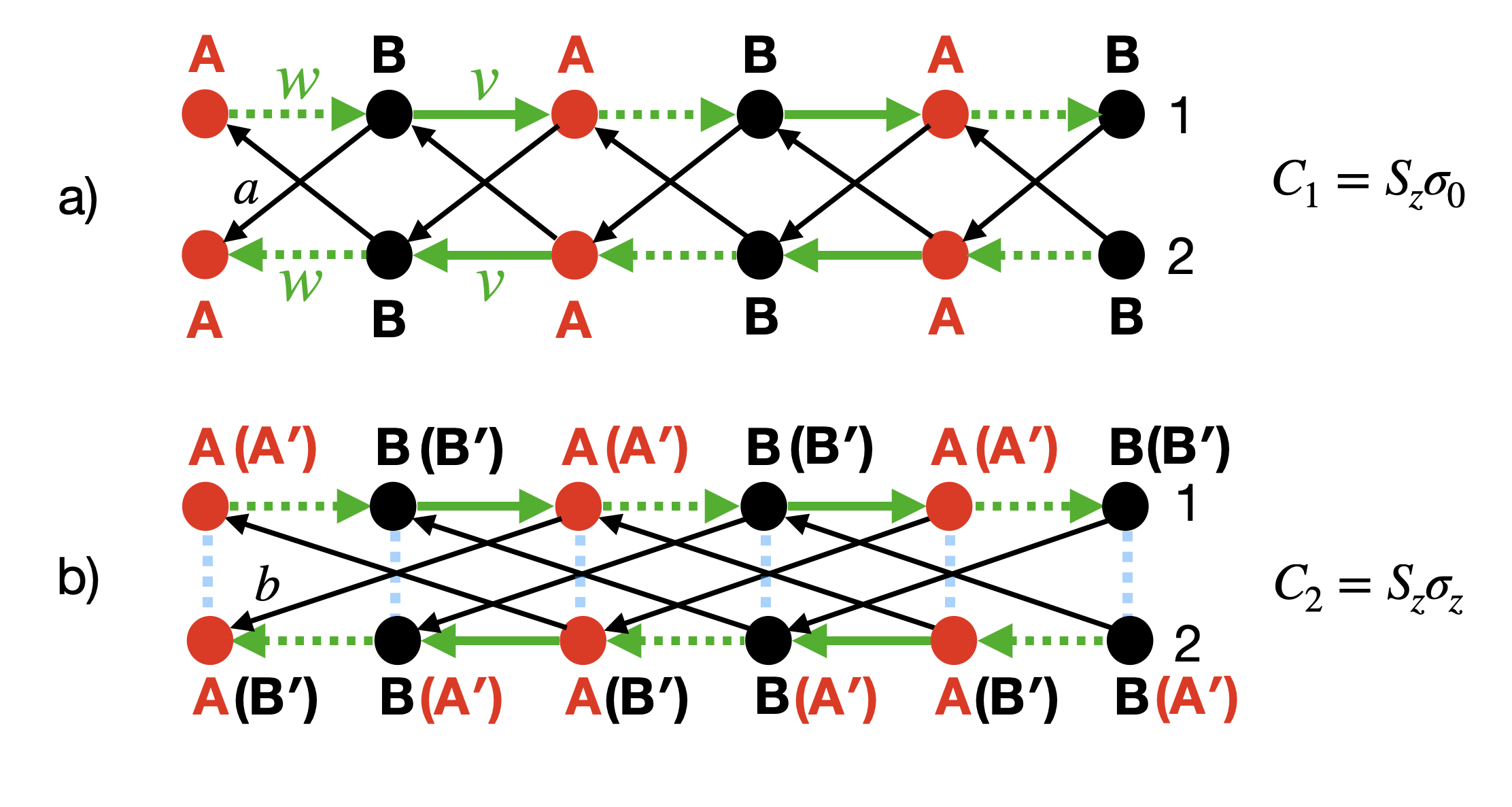}
\caption{Two ways of constructing AIII topological class by coupling SSH chains.  Figure a) illustrates a system with chiral symmetry $C_1=S_z\sigma_0$ and Figure b) illustrates the coupling structure corresponding to $C_2=S_z\sigma_z$.  Labels A and B denote the `natural' choice of the sublattices, according to the chiral symmetry $C_1$ and $\text{A}', \text{B}'$ denote the sublattices defined according to $C_2$. Corresponding projectors onto those sublattices are defined in (\ref{sublatticeprojection}). }
\label{Coupling}
\end{figure}

As there are two possible choices for chiral symmetry, there are also two possible types of coupling terms.  One of them preserve chiral symmetry $C_1=S_z\sigma_0$ and the other preserves $C_2=S_z\sigma_z$. These two configurations are illustrated in Fig. \ref{Coupling}.  By rephrasing the chiral symmetry in terms of the sublattice symmetry \eqref{sublatticeprojection} the construction physically corresponds to connecting atoms belonging to different sublattices,  but zero coupling between the same sublattice.  In Fig.  \ref{Coupling} we illustrate two types of sublattices A/B and $\text{A}'$/$\text{B}'$ corresponding to the chiral symmetries $C_1$ and $C_2$.  The difference between the two chiral operators is a relabelling of the sublattice on the lower chain.

Without any further symmetry constraints on the inter-chain coupling, the two pictures illustrate two inequivalent topological models belonging to the AIII class.
Applying further constraints to the interchain coupling allows us to fulfil symmetry requirements for models of CI, CII, DIII, and BDI classes. In the next subsections we 
construct microscopic models that represent these classes.

\subsection{Time-reversal and particle-hole symmetry}

We have discussed at length the chiral symmetry (and ambiguity therein) of two uncoupled chains, Hamiltonian \eqref{block_H0}.  Let us now turn to the issue of time-reversal symmetry.

We have very specifically formulated our SSH model with complex hopping terms, so that $\hat{H}_\text{SSH} \ne \hat{H}_\text{SSH}^*$ -- meaning that each individual chain does not possess time-reversal symmetry.  However in our construction, the second chain is the time-reversal counterpart of the first.  This means that the overall model of two chains should obey time-reversal symmetry.  From the basic definition, Eq.~\eqref{symmetries}, the time reversal symmetry operator is a unitary operator times the complex conjugate operator that commutes with the Hamiltonian \eqref{block_H0}.  It turns out that there are two different operators satisfying this that one could write -- one squaring to $+1$ and the other squaring to $-1$.
\begin{equation}
\label{time-reversal}
\begin{cases}
 T^2=+1 : \hspace{0.3cm} T_+= U^{+}_T \mathcal{K}, U^{+}_T=S_0 \sigma_x \\
 T^2=-1 : \hspace{0.3cm} T_-= U^{-}_T \mathcal{K}, U^{-}_T=iS_0 \sigma_y
\end{cases} 
\end{equation}
The structure of these can be interpreted physically -- in order to satisfy time reversal, one needs to swap the chains i.e. the time reversal operator must be proportional to $\sigma_x$ or $\sigma_y$.  It is however somewhat perplexing that there is two distinct time reversal operators that are both symmetries of the decoupled two-chain system.  Like the choice of chiral operators, this is due to the block diagonal structure and hence additional unitary symmetry the decoupled model exhibits.  As soon as one breaks this symmetry by an inter-chain coupling, this ambiguity will evaporate.

We can play the same game with particle-hole symmetry operators -- from Eq.~\eqref{symmetries}, this should be a unitary operator times the complex conjugate operator that anti-commutes with the Hamiltonian \eqref{block_H0}.  Again, calculation reveals two such operators:
\begin{equation}
\label{particle-hole}
\begin{cases}
 P^2=+1 : \hspace{0.3cm} P_+= i U^{+}_P \mathcal{K}, U^{+}_P=S_z \sigma_x \\
P^2=-1 : \hspace{0.3cm} P_-= U^{-}_P \mathcal{K}, U^{-}_P=-iS_z \sigma_y,
\end{cases}
\end{equation}

While we have previously discussed the chiral (sublattice) symmetry operator in isolation, let us now recall that it originally was defined as the combination of particle hole and time reversal -- $C= P\cdot T$.  By taking different choices of $T$ and $P$ symmetries, we find an alternative way to reproduce the two different $C$ operators we previously discussed:
\begin{equation}
\label{symmetries_representation}
\begin{cases}
C_1 = P_+ T_+ = P_- T_- = S_z\sigma_0 & \text{BDI, CII classes}\\
C_2 = P_- T_+ = P_+ T_- = S_z\sigma_z & \text{CI, DIII classes}.
\end{cases}
\end{equation}
While in the case with no additional symmetries, the choice of chiral operator was motivated only by how the sublattices are divided, we see now that this has implications on what other symmetries may or may not be present.  We will discuss this further as we discuss each of the universality classes in turn.

We can also see from this that the decoupled model may be classified in any one of four \textit{different} universality classes -- BDI, CII, CI or DIII.  While this ambiguity is indeed explained by the extra symmetry that will be broken when a coupling is added, it is particularly surprising as the topological classification of each of these cases is not the same -- BDI and CII are $\textbf{Z}$ topological insulators; DIII is a $\textbf{Z}_2$ topological insulator, and CI is topologically trivial.  We will resolve this seeming paradox in the next sections as we discuss some concrete models.

To deeper understand and resolve these ambiguities, we now construct models by coupling the chains.   In the presence of inter-chain coupling the Hamiltonian of two chains (\ref{block_H0}) becomes
\begin{align}
\label{block_H}
\hat{H}=\begin{pmatrix}
\hat{h}_{\text{SSH}} & \hat{W} \\
\hat{W}^{\dagger} & \hat{h}^*_{\text{SSH}}
\end{pmatrix},
\end{align} 
The off-diagonal block $\hat{W}$ describes the coupling between the chains.  Note that for physical transparency, we are using a basis $c^{\text{T}}_n=\left\{c_{A ,1,n}, c_{B,1,n},c_{A,2,n},c_{B,2,n} \right\}$ corresponding to a block structure in chain space; not in sublattice space.  Reordering the basis would give the block off-diagonal structure \eqref{chiral_ham2} that makes explicit the chiral symmetry.

Our goal is to build the coupling $\hat{W}$ that is compatible with firstly chiral symmetry, but also time-reversal and particle-hole symmetries in order to construct models in different chiral universality classes.

 \subsection{Real space realizations of one-dimensional topological insulators}
 
 \subsubsection*{Classes BDI and CII}
 
 Let us first consider the case where the chiral symmetry operator is $C_1=S_z\sigma_0$ -- meaning that the chains are coupled according to Fig. \ref{Coupling}a.  According to \eqref{symmetries_representation}, this includes models in the BDI and CII universality classes.  The minimal lattice model for this would have all couplings identical (general model is described in the Appendix \ref{AppendixC}):
\begin{equation}
 \hat{H}_1 = \hat{H}_0+ \hat{V}_1
 \label{H1}
 \end{equation}
 where $\hat{H}_0$ is given by \eqref{block_H0} and the coupling between chains is
  \begin{align}
 \label{BDI_CII}
 \hat{V}_1 &= a \sum_n \left( c^\dagger_{A,1,n} c_{A,B,n} + c^\dagger_{A,2,n} c_{B,1,n} \right. \nonumber \\
 & \left. + c^\dagger_{B,1,n} c_{A,2,n+1} 
 + c^\dagger_{B,2,n} c_{A,1,n+1}    \right) + h.c.
 \end{align}
 The strength of the coupling is given by the parameter $a$.  It is straightforwards to show that if $a$ is real, then this Hamiltonian is symmetric under $T_+$ and $P_+$ meaning that the model falls in the BDI universality class, while if $a$ is imaginary, the Hamiltonian is symmetric under $T_-$ and $P_-$, so the model is in the CII class.  In the case where $a$ is complex (i.e. neither real nor imaginary), the model has no additional symmetry beyond the chiral symmetry, and hence is in universality class AIII.
 
 Let us suppose that the coupling between chains $|a|$ is small -- specifically small compared to $|v|-|w|$ so that the gap does not close.  Then the winding number 
 is given by the sum of the winding numbers of the decoupled chains $\nu=\nu_1+\nu_2$.
 Therefore, in this limit the model (\ref{BDI_CII}) has the following phases: 
\begin{align}
\label{winding_BDI}
\nu_{\text{tot}}=\begin{cases}
2, \hspace{0.5cm} \text{if}  \hspace{0.5cm} |w/v|<1 \\
0, \hspace{0.5cm}\text{if} \hspace{0.5cm} |w/v|>1 .  
\end{cases}
\end{align}
This can be verified by a direct calculation of $\nu_\text{tot}$ through Eq.~\eqref{phase_detq}.

Therefore, for weakly coupled chains there are two phases possible: topologically trivial and $\nu=2$.  Notice that this calculation of winding number is independent of the complex phase of the coupling $a$ -- i.e. it is independent of whether the system lies in universality class AIII, BDI or CII.  In other words, the gapped phases in these three universality classes are all topologically equivalent in one dimension.  One can take a path $a=|a|e^{i\theta}$ from $\theta=0$ to $\theta=\pi/2$ and the zero-mode edge states will remain for the entire path.

There is a slight subtlety here, that the $T_-$ symmetry leads to all states have a Kramers partner.  This means that the winding number in CII must be even -- the classification is $2\mathbf{Z}$ rather than $\mathbf{Z}$.  This means that a model with an odd winding number may never be in the CII class -- however for any model with an even winding number, it may be adiabatically deformed to the CII class without closing the gap and without affecting the edge states.

Let us now turn to the edge states -- to be concrete we will focus on a left edge of a semi-infinite chain, the calculation for the right edge is analogous.  In appendix \ref{appendix_edge} we show that the left edge states for model \eqref{H1} is localised on the A sites only (i.e. the amplitude on the B sites is zero) and may be written in the basis $\begin{pmatrix} A1 & A2 \end{pmatrix}$ as
\begin{equation}
\underline{\psi}_{\pm}(n) = \lambda_1^n \underline{u}_1 \pm \lambda_2^n \underline{u}_2
\label{eq:edge1}
\end{equation}
where $\lambda_{1,2}$ are the complex eigenvalues of a transfer matrix and $\underline{u}_{1,2}$ are the corresponding eigenvectors.  The edge states are normalisable when $|\lambda_{1,2}|<1$ which corresponds to the topological phase.  As the two edge states are degenerate one can take any linear combination, however our choice of $\underline{\psi}_\pm$ is useful to demonstrate some properties of these edge states.

\begin{figure}
\includegraphics[scale=0.32]{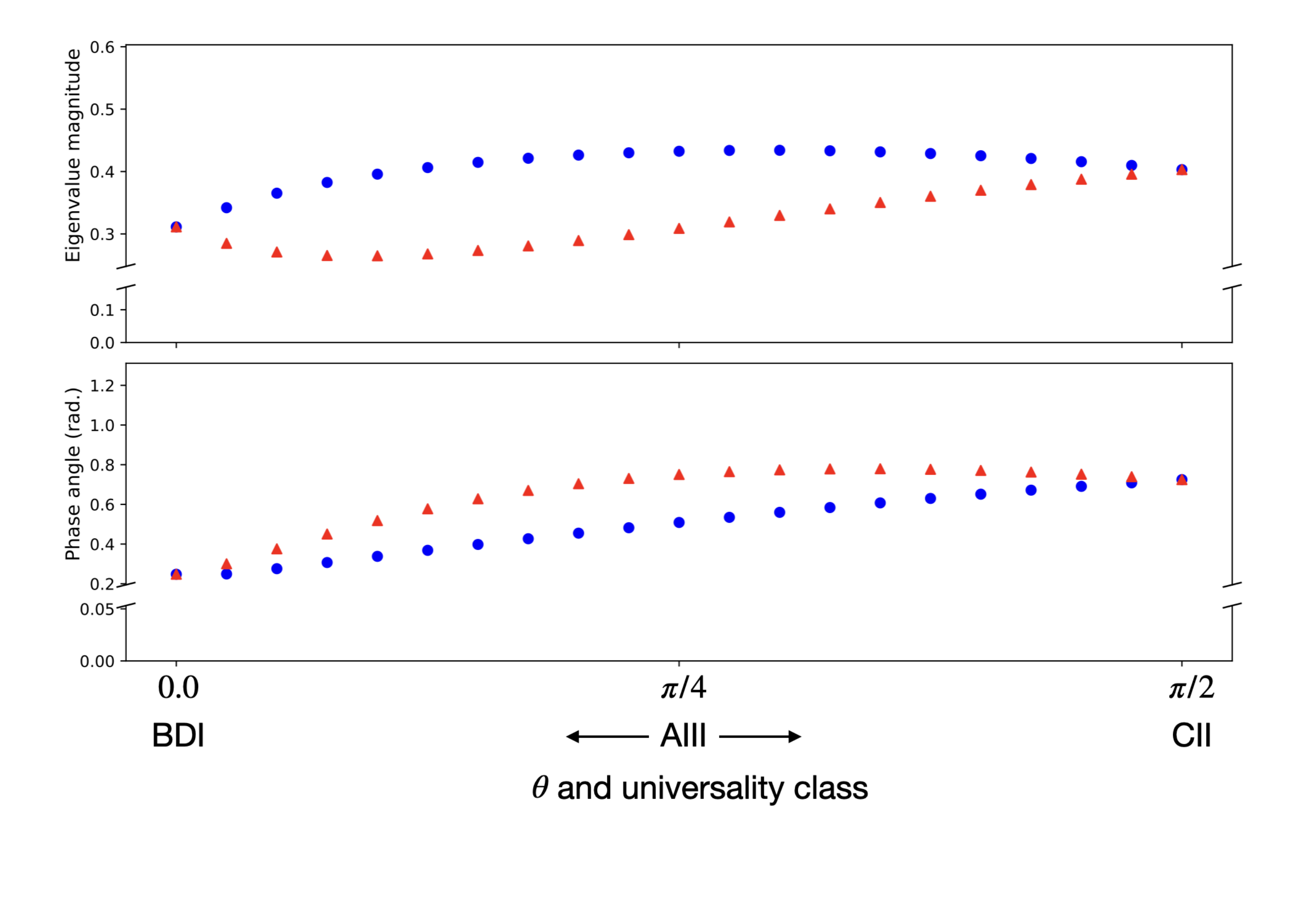}
\caption{Amplitudes and absolute value of phases for the eigenvalues of transfer matrix $\lambda_1$ and $\lambda_2$ defined in (\ref{eq:edge1}).  The phase $\theta$ of inter-chain hopping parametrizes the path between two topological classes BDI and CII. }
\label{figure_edge_states}
\end{figure}

For generic complex $a$ (the AIII universality class), there is no particular relationship between $\lambda_1$ and $\lambda_2$ -- both the amplitude and phase of these complex numbers will be different, meaning there are two edge states with (slightly) different decay lengths and different wavevectors for oscillation.  However if one goes to one of the points with time reversal symmetry, i.e. $a$ is real corresponding to the BDI class or $a$ is imaginary corresponding to the CII class, we see that $\lambda_1=\lambda_2^*$.  This is shown in Fig.~\ref{figure_edge_states}.
Ultimately, this relationship between the constants $\lambda$ illustrates something about time reversal symmetry.  For the BDI class, we see that
\begin{equation}
T \underline{\psi}_\pm = \pm \underline{\psi}_\pm
\end{equation}
i.e. each edge state is itself time reversal symmetry.  For the CII class however, we see that
\begin{equation}
T \underline{\psi}_\pm = \pm \underline{\psi}_\mp
\end{equation}
which shows that the two edge states form a Kramers doublet.  We emphasise that the result \eqref{eq:edge1} for the edge states is valid for any complex $a$ -- i.e. the edge states evolve continuously as one takes a path between universality classes from BDI through AIII to CII.  The only difference one finds in the edge states between these classes is the time-reversal properties -- if the universality class has time reversal symmetry, so do the edge states.  This is summarised in table \ref{tab:edge1}

\begin{table}
\begin{tabular}{c|p{5.8cm}}
Universality class & property of edge states \\ \hline
BDI & Edge states are time reversal symmetric.  N.B. there can be a phase factor $\pm 1$ so one may need the correct linear combination of degenerate edge states to see this. \\
AIII & Edge states have no symmetry properties \\
CII & Edge states from Kramers pairs which are time reversal partners of each other.
\end{tabular}
\caption{A table summarising the properties of the edge states}
\label{tab:edge1}
\end{table}

Let us now briefly consider what may happen if the coupling between the chains is no longer weak.  By increasing the coupling strength $a$ in the model (\ref{BDI_CII}) one may close the gap.  At this point the system undergoes a phase transition to a different topological phase.  For instance,  in the case when the coupling parameter $a$ is real (BDI class),  the gap closes if $a = \pm (|v|-|w|)/2$,  so when the coupling strength becomes of the order of the gap in an uncoupled system.  At this transition point the winding number changes by 1. 
In the case when the parameter $a$ is imaginary,  which corresponds to the CII class,  the winding number is determined only by the ratio $|w/v|$.  As in the case of weakly coupled chains,  (\ref{winding_BDI}) the winding number $\nu =2$ if  $|w/v| <1$ and $\nu=0$ if $|w/v|>1$.  
 
 \subsubsection*{Classes DIII and CI}
Let us now construct the classes that correspond to the second type of coupling, see Fig. \ref{Coupling}b, corresponding to a chiral symmetry operator $C_2=S_z\sigma_z$.  In this case, direct couplings between the two chains on the same site are allowed, but it turns out to be more useful for us to consider this coupling to be zero and take a minimal model to include the interchain coupling $b$ between next-nearest neighbors.  This model may be written
\begin{equation}
 \hat{H}_2 = \hat{H}_0+ \hat{V}_2\nonumber \\ 
 \end{equation}
 where $\hat{H}_0$ is given by \eqref{block_H0} and the coupling between chains is
  \begin{align}
 \label{DIII_CI}
 \hat{V}_2 &= b \sum_n \left( c^\dagger_{B,1,n} c_{B,2,n+1} + c^\dagger_{B,2,n} c_{B,1,n+1} \right. \nonumber \\
 & \left. + c^\dagger_{A,1,n} c_{A,2,n+1} 
 + c^\dagger_{A,2,n} c_{A,1,n+1}    \right) + h.c.
 \end{align}
If the inter-chain coupling $b$ is imaginary, then this model is symmetric under $T_-$ and $P_+$, leading to the DIII classification; if $b$ is real then the model is symmetry under $T_+$ and $P_-$ leading to the CI classification, while if $b$ is any other complex number, there are no additional symmetries and the classification is AIII. Note that the Hamiltonian (\ref{DIII_CI}) is not the most general, more terms compatible with the symmetries can be added, see Appendix \ref{AppendixC}. 

As discussed in the general introduction, if $|b|$ is small, the winding number for this type of coupling is $\nu_\text{tot}=\nu_1-\nu_2=0$ -- again, a direct calculation confirms this.

While the winding number is necessarily zero for this choice of couplings, in the case of DIII where we have $T_-$ symmetry, we can define a $\mathbb{Z}_2$ invariant.    This is given by parity of the winding number of one of the Kramers partners \cite{Budich2013}. As weakly coupled chains are topologically equivalent to a pair of non-coupled chains with Hamiltonians related by time-reversal symmetry, their eigenstates are Kramers partners, thus the $\mathbb{Z}_2$ index is determined by a winding number of one of the chains, and is given by:  
\begin{align}
\label{Z2_DIII}
p=(-1)^{\nu_{\text{SSH}}}, \nonumber \\
\nu_{\text{SSH}}=\begin{cases}
-1, \hspace{0.5cm} \text{if}  \hspace{0.5cm} |w/v|<1  \\
1, \hspace{0.8cm}\text{if} \hspace{0.5cm} |w/v|>1.  
\end{cases}
\end{align}
The case with $p=-1$ hosts edge states and thus corresponds to a topologically non-trivial phase.
 The phase with $p=1$ does not have edge modes and therefore is topologically trivial. 
When the inter-chain coupling amplitude $b$ becomes large,  i.e.  of the order of the  gap in uncoupled system, the gap might be closed.  In particular,  when one starts with the topologically non-trivial phase, $|\omega|<|v|$,  by increasing the parameter $b$ one drives the system to topologically trivial phase.  In the trivial phase when $|\omega|>|v|$ the gap does not close as one tunes $b$. 

Note that in contrast to the BDI, CII and AIII cases, the existence of zero energy edge states of the model is not protected by chiral symmetry, as the winding number vanishes.  As we will discuss below in the broader context of generic models,  other symmetries (time-reversal $T_-$ and particle-hole $P_+$) are needed to protect the degeneracy of the edge states.  We confirm that statement by explicitly computing the edge states of the DIII model in Appendix \ref{appendix_edge}.  We find the following properties of the edge states:
\begin{itemize}
\item In the basis $\begin{pmatrix} A'1 & A'2 \end{pmatrix}$, one of the edge states may be written as
\begin{equation}
\underline{\psi}_{\pm}(n) = \lambda_1^n \underline{\psi}_{0,1} - \lambda_2^n \underline{\psi}_{0,2}
\label{eq:edge1}
\end{equation}
where $\lambda_{1,2}=e^{\delta_{1,2}}$ are certain (complex) eigenvalues of a transfer matrix with magnitude less than 1 given by Eq.~\eqref{delta_tr}, and $\underline{\psi}_{0,1(2)}$ are the corresponding eigenvectors, given in Eq.~\eqref{solutions_CI_spinors}.
\item This edge state is localised on the $A'$ sub-lattice.  The other edge state is the time-reversal (Kramers) partner, which in this case is localised on the $B'$ sub-lattice.  This is in sharp contrast to the case of CII where both Kramers partners were localised on the $A$ sub-lattice.  We will discuss this difference in detail in the next section.
\item If one perturbs away slightly from the DIII point in phase space (i.e. perturb away from the case where the parameter $b$ in Hamiltonian \eqref{DIII_CI} is purely imaginary), one can no longer find normalisable zero-energy states satisfying the boundary conditions -- in other words, the topologically protected edge states require the extra symmetries beyond chiral.  It is worth noting in passing that the exponentially decaying solutions still exist, just they do not satisfy the simple boundary conditions implied by the end of a chain without splitting a unit cell.
\end{itemize}

Having now seen the properties of the edge state of a specific DIII model in contrast to specific models in the classes AIII, BDI or CII, we go on to show that the majority of these properties are not model-specific and are in fact general for all models of these classes.

%%%%%%%%%%%%%%

\section{General properties of the edge states}
\label{section3}
\subsection{Chiral symmetric models}
It is well known that the edge states of the SSH model are localised on the A sublattice of the left edge (and the B sublattice on the right edge).  We have shown that this continues to be true for our model with the first type of chiral symmetry $C_1$, which has non-zero winding number (classes AIII, BDI and CII).   For the second type of chiral symmetry $C_2$ the story is slightly more complicated -- each edge state is still confined to only one sublattice, so long as one defines the sublattice in accordance with the chiral symmetry,  i.e through Eq.~\eqref{sublatticeprojection}.  In this case however, one of the left edge states was on sublattice $\text{A}'$ while the other was on sublattice $\text{B}'$.  This happens for class DIII, where the winding number is zero.    The difference between this case and the first is that in the first, the left edge states belong only to sites on the A sublattice, while in the second case one of the left edge states is localised on sublattice $\text{A}'$ while the other left edge state is on sublattice $\text{B}'$.

It is quite simple to prove that any 1D model with chiral symmetry must have the property that the edge states are localised only on one of sublattices.  From Eq.~\eqref{sublatticeprojection}, we can write the chiral symmetry in terms of the sublattice projectors
\begin{equation}
U_C = P_A - P_B.
\label{UCPAPB}
\end{equation}
We also know that $U_C$ acting on an eigenstate of the Hamiltonian will give a state with negative the energy $U_C | E \rangle = | - E \rangle$, this is a direct consequence of the anticommutation between $U_C$ and the Hamiltonian.   Hence $U_C$ acting on a state within the zero-energy subspace (the space of edge states in a gapped model) will remain within this subspace.  From the structure of $U_C$ above,  Eq.~\eqref{UCPAPB}, we can also see that $U_C$ acting on an edge state on the left edge must create another edge state on the left edge and similarly for the right edge.

Let us start with a model with one edge state on the left edge (e.g. the SSH model).  This state must therefore be an eigenstate of $U_C$.  By the structure $U_C = P_A - P_B$, we see that the only way this can happen is if $P_A |\psi \rangle =0$ or $P_B |\psi \rangle =0$, i.e. the state must be localised on one sublattice.  Which one will depend on details of the model and the edge.

We can extend this to models with more than one edge state on the left edge.  In this case, one can find linear combinations of these states such that they are all eigenstates of $U_C$ because the left edge states are a closed subspace under the chiral symmetry operator as we have just discussed.  One can then apply the same logic to each edge state -- each one must be localised on either the A or B sublattice.

Furthermore we can prove that the edge states protected by chiral symmetry are localised on the same sublattice. 
Consider two edge states $|\Psi_1\rangle$ and $|\Psi_2 \rangle$ such that both of them are eigenstates of the chiral symmetry operator,  $C|\Psi_{1,2}\rangle=\alpha_{1,2} |\Psi_{1,2} \rangle$,  where the eigenvalues $\alpha_{1,2}$ are each $\pm 1$ (because $U_C^2=1$).  We can add some perturbation $\hat{V}$ that acts within this subspace and preserves chiral symmetry,  i.e. $\{\hat{V}, C\}=0$.  In this case the matrix element  $\langle \Psi_{1}| \hat{V}|\Psi_{2}\rangle= \alpha_1\alpha_2 \langle \Psi_{1}| C^{\dagger} \hat{V} C |\Psi_{2}\rangle= -\alpha_1\alpha_2 \langle \Psi_{1}| \hat{V} |\Psi_{2}\rangle$. This matrix element vanishes if the edge states are the eigenstates of the chiral symmetry operator with the same eigenvalue, i.e.  if $\alpha_1=\alpha_2$.  Therefore if the states are topologically protected, they \textbf{must} be localised on the same sublattice. 

In the case of symmetry class DIII, we can therefore see clearly that the edge states are not protected by chiral symmetry alone -- the addition of a weak perturbation respecting chiral symmetry (but breaking the other ones) can hybridise the edge states.  This is consistent with the winding number of zero.

\subsection*{Particle-hole symmetry}
Chiral symmetry and particle-hole symmetry have a similar property.  They both anticommute with the Hamiltonian,  however an important difference between them is that chiral symmetry is represented by a unitary operator and particle-hole is anti-unitary.  
This is manifested in the fact that the classes with particle-hole symmetry only (D and C) have different classification rather than the class AIII with chiral symmetry. In particular, class D obeys $\mathbb{Z}_2$ classification and C is topologically trivial.  Here we would like to explain those differences. 

First we focus on the case of $P_+$ symmetry.  This symmetry may protect a single zero-mode edge state.  To prove this,  we rely on the fact that like for chiral symmetry, $P_+|E \rangle \propto |-E \rangle$; hence a single zero-energy state remains pinned to zero energy as long as $P_+$ remains a symmetry.

Now consider the case of two edge states, and recall that chiral symmetry would protect these so long as they are both localised on the same sublattice.  We now prove this is not the case for the $P_+$ symmetry.
 
As a basis in this two-dimensional space we can choose an orthogonalised pair of eigenstates of $P_+$ (note that eigenstates of an anti-unitary operator are not necessarily orthogonal, but in this case they are).  One can easily show that the particle-hole symmetry operator in this basis has the following representation: 
\begin{align}
P_+ = \begin{pmatrix}
\alpha_+ & 0 \\0 & \alpha_- 
\end{pmatrix}K 
\end{align}
where $\alpha_{\pm}$ are the eigenvalues of $P_+$.  While $P_+^2=1$, the fact that $P_+$ includes the complex conjugation operator means that these eigenvalues satisfy only the requirement $|\alpha_\pm|=1$.   From $P_+^2=1$ it follows that $|\alpha_{\pm}|=1$.  Next, we consider some generic perturbation that respects the $P_+$ symmetry: 
\begin{align}
\label{PH_condition}
P_+\hat{V}P_{+}=-\hat{V}.
\end{align}
Let us suppose that this operator can hybridise the two edge states, so we write it as
\begin{align}
\label{V_pert}
V=\begin{pmatrix}
0 & b \\ b^* & 0
\end{pmatrix}.
\end{align}
By doing the matrix multiplication,
 \begin{align}
P_+\hat{V}P_{+}= \begin{pmatrix}
0&  b^* \alpha^*_- \alpha_+ \\ b \alpha_- \alpha^*_+ & 0
\end{pmatrix},
\end{align}
we see that the particle-hole symmetry condition is satisfied if
 \begin{align}
 \label{V_condition}
b^* \alpha^*_- \alpha_+ =-b 
\end{align}
From here one can determine the phase of $b$ if we denote $b = |b|e^{i\phi_b}, \alpha_\pm = e^{i\phi_\pm}$: 
\begin{align}
 \phi_b=(\phi_+-\phi_--\pi)/2
\end{align}
Hence we have shown that for any two edge states localised on the left edge at zero energy, we can find an operator that respects particle hole symmetry that hybridises these two edge states.  In other words,  these edge states can not be topologically protected by the $P_+$ symmetry alone.  It is easy to extend this argument to show that if there are an odd number of edge states, one of them will not be hybridised, while an even number will have no protection.  This explains the $\mathbf{Z}_2$ classification of topological insulators in the D universality class in 1D.

Let us note that one can follow a similar argument for chiral symmetry -- however in this case we do not get the complex conjugate, and hence the case $\alpha_+\alpha_-=1$ would mean $b=0$, i.e. there is no operator with the correct symmetry that one can write that would hybridise the two edge states.
 
Now we turn to the class C with $P_-$.  An equivalent argument to Kramers theorem tells us that in this case the edge states must come in pairs with energies $\pm E$.  A minor extension of the above argument shows that one can always write a small perturbation with the $P_-$ symmetry that hybridises a pair of edge states with $E=0$.  Hence there can never be any topologically protected edge states in this class.

\subsection{The case of DIII}

The class DIII has chiral symmetry, but zero winding number.  We have shown in this case that there is a pair of zero-energy edge states on the left edge, one localised on the A' sublattice and the other on the B' sublattice.  We know that a generic perturbation respecting only chiral symmetry will hybridise these.

This class also has the $P_+$ symmetry, but we have just shown that this symmetry alone will protect only one edge state; not a pair of them.

Hence the pair of edge states in the class DIII require all symmetries to remain unhybridised and pinned to zero-energy.  One way to think of this is that the $T_-$ symmetry enforces states to come in Kramer's pairs with the same energy.  The $P_+$ symmetry requires states to come in pairs with $\pm E$.  With a single pair, the only way to satisfy both these conditions is $E=0$ -- but we emphasise that both time reversal and particle-hole symmetry are crucial in this case.

\section{Construction of $N$-chain chiral symmetric model}\label{section4}
One can iteratively extend the scheme of constructing topological models by coupling two chains to an arbitrary number of pairwise coupled chains.  This allows for the construction of a generic $2N$ band AIII model of 1D topological insulators with a given winding number. We consider a set of $N$ chains.  Next, we want to build all possible inequivalent chiral symmetry operators. 

One can choose $l=N/2+1$ inequivalent chiral symmetry operators if $N$ is even and $l=(N+1)/2$ if $N$ is odd,  as follows: 
\begin{align}
\label{S_i}
C_i= \text{M}_i \otimes S_z, \hspace{0.5cm} i=1,2...l. \nonumber\\
\text{M}_i= \begin{pmatrix} 
- \text{I}_{i} & 0 \\ 
0 & \text{I}_{N-i} 
\end{pmatrix},
\end{align}
here $S_z$ acts in a space of sublattices A and B and $\text{M}_i$ acts in the chain basis.  $\text{I}_{i}$ denotes an identity matrix of the size $i \cross i$. In addition, an arbitrary permutation of chains generates a valid symmetry operator $M_i$, that corresponds to the permutations of the elements $1$ and $-1$ on the diagonal. This corresponds to additional $m={{N}\choose{i}}$ non-equivalent ways of coupling for a fixed $i$. We will therefore assign an additional index to the chiral symmetry operator: $C^k_i$, where $k=1..m$. 

The corresponding winding number in case of generic weak coupling that is compatible with chiral symmetry $C^k_i$ is given by: 
\begin{align}
\label{winding_N}
\nu^i= \sum\limits_{j\in M_{+}} \nu_j -\sum\limits_{j\in M_{-}} \nu_{j} \\ 
\nu^i \in [0,\text{max}[i, N-i]]. \nonumber
\end{align}
Thus, the winding number is the difference between the winding number of chains with the chiral operators $S_z$ and with operator $-S_z$. Thus, the largest winding number of a set of $N$ coupled chains is $\nu=N$ in the case of $i=0$ and the minimal is $\nu=0$. Here we denoted by $M_{+}$ a set of chains with a chiral symmetry $C=S_z$ and by $M_{-}$ set of chains with $C=-S_z$. 
To illustrate these general statements on a simple example, we focus on the case of $N=3$ chains. 
\subsection*{Example: Three coupled chains}
\begin{figure}
\includegraphics[scale=0.4]{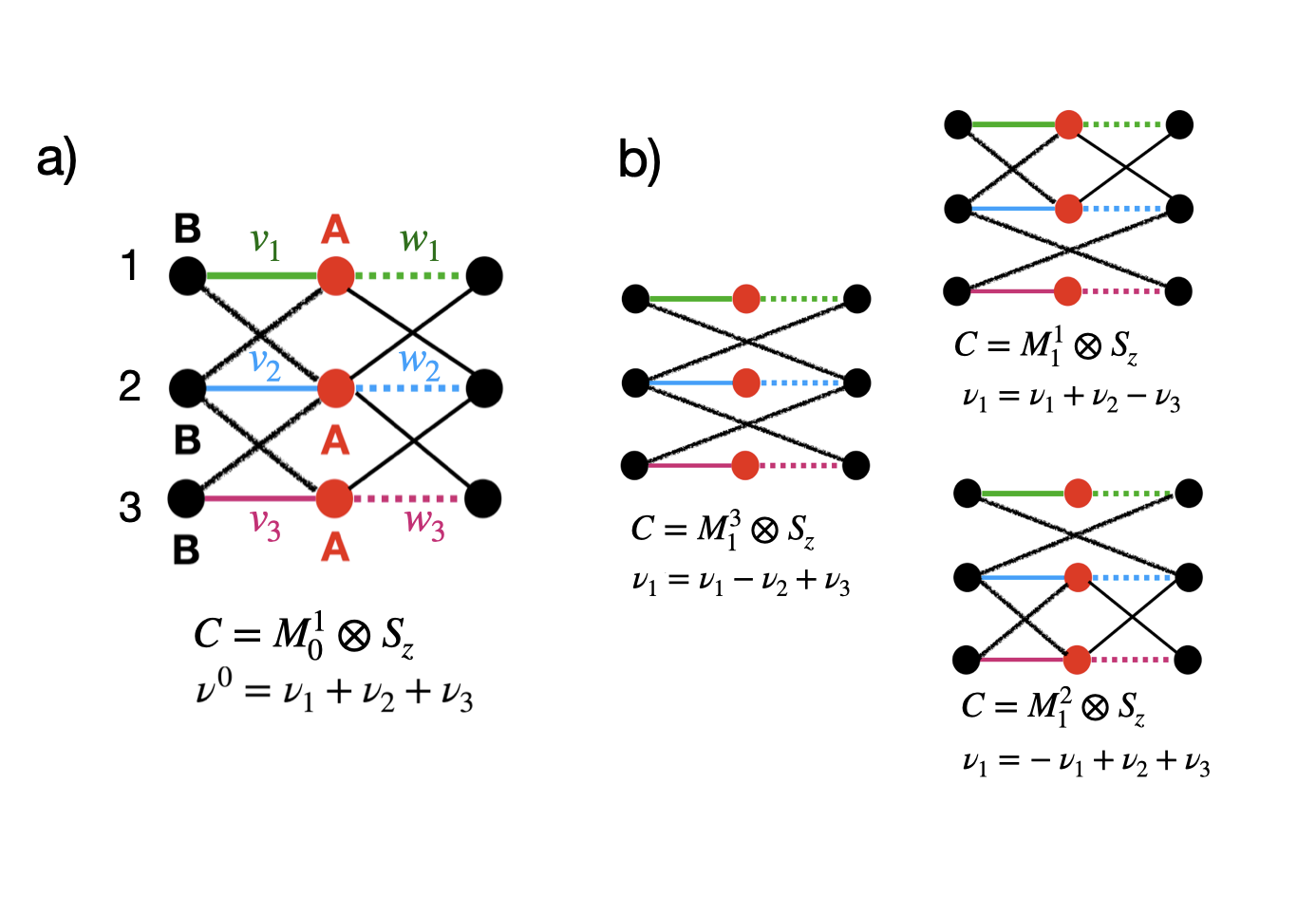}
\caption{Possible coupling structure of three SSH chains and corresponding winding numbers for each type of coupling, according to (\ref{3chains}) and (\ref{winding_N}). }
\label{3chainspic}
\end{figure}

As a first step, we explicitly write all possible chiral symmetry operators determined by the matrix $\text{M}_i$ according to (\ref{S_i}): 
\begin{small}
\begin{align}
\label{3chains}
\nonumber
M^{1}_0=\begin{pmatrix} 
1 & 0 &0 \\ 
0 & 1 & 0 \\
0 & 0 & 1 \\
\end{pmatrix}, \nonumber \\ 
M^{1}_1=\begin{pmatrix} 
1 & 0 &0 \\ 
0 & 1 & 0 \\
0 & 0 & -1 \\
\end{pmatrix},
M^{2}_1=\begin{pmatrix} 
-1 & 0 &0 \\ 
0 & 1 & 0 \\
0 & 0 & 1 \\
\end{pmatrix}, 
 M^{3}_1=\begin{pmatrix} 
1 & 0 &0 \\ 
0 & -1 & 0 \\
0 & 0 & 1 \\
\end{pmatrix}.  
\end{align}
\end{small}
We schematically illustrate corresponding coupling structures in the Fig. \ref{3chainspic} for three coupled SSH chains with arbitrary in-chain hopping amplitudes $v_i$ and $\omega_i$. We evaluate corresponding winding numbers for each type of coupling according to (\ref{winding_N}): 
\begin{align}
\label{winding_3chaiins}
\nu^0=\nu_1+\nu_2+\nu_3, \hspace{0.5cm} \nu^0 \in [0,3] \nonumber \\ 
\nu^1= \begin{cases}
\nu_1+\nu_2-\nu_3 \\
-\nu_1+\nu_2+\nu_3 \\
\nu_1-\nu_2+\nu_3  
\end{cases} \hspace{0.5cm} \nu^1 \in [0,2].
\end{align}

Figure  \ref{3chainspic}a corresponds to the chiral symmetry $C_0^1=M^{1}_0 \otimes S_z$ and the winding number is given by $\nu^0$ (see Eq.~\ref{winding_3chaiins}). Figure \ref{3chainspic}b illustrates the chain models with chiral symmetries $C_1^1=M^{1}_1 \otimes S_z$, $C_1^2=M^{2}_1 \otimes S_z$ and $C_1^3=M^{3}_1 \otimes S_z$. The winding number of these models is given by $\nu^1$ in (\ref{winding_3chaiins}). 
Systems with a larger number of chains can be constructed iteratively. One can formulate a general rule for coupling two neighboring chains: if two neighboring chains are assigned different signs in the operator $C^k_i$, they should be coupled according to Figure \ref{Coupling}b and if they have the same sign, one should couple them according to Figure \ref{Coupling}a. Thus we demonstrated how to construct a generic chiral symmetric multi-chain system and evaluated its winding number in the case when the coupling is weak. 
\section{Experimental realization}
So far we have been focusing on purely theoretical models. 
To observe the above-studied effects one needs to realize these models in experiments.
Right now, the SSH model has been studied experimentally with ultracold atoms \cite{Bloch2013,Leder2016}.
However, it remains to connect a possible experimental realization with microscopic models described in our paper 
for all choices of symmetry classes. It seems to be feasible in cold atomic settings. 

Concretely, the coupled SSH chains can be viewed as models for spinful fermions. In that case, coupling terms correspond to spin-orbit interaction and Zeeman terms, with staggered amplitudes. One can realize these types of terms within cold atomic setups, by extending the existing scheme for SSH potential by taking atoms with additional internal degrees of freedom. The staggered magnetic field can be realized by creating an inhomogeneous magnetic field with the period half of the lattice constant and spin-orbit terms emerge when one couples internal degrees of freedom by additional lasers \cite{Campbell2011, Lin2011, Wang2012, Cheuk2012}. Moreover, also the phases of the parameters in the Hamiltonian can be controlled independently \cite{Aidelsburger2013, Aidelsburger2011}. 

\section{Discussion and conclusion}

 We studied one-dimensional non-interacting topological insulators  with chiral symmetry.  
 We build our models from coupled one-dimensional chains. Each of the uncoupled models is described by a two-band Hamiltonian with two sublattices $A$ and $B$ and is characterized by an integer topological invariant -- the winding number. 
 We showed, that switching the labels of sublattices $A \leftrightarrow B$ switches the sign of the winding number, this may also be thought of as switching the sign of the protecting chiral symmetry operator.  This implies, that there are multiple ways of constructing the coupled system with chiral symmetry, that correspond to inequivalent types of coupling between the chains. By choosing the specific type of coupling, one removes the freedom of relabeling of the sublattices in the individual chains (although it remains overall in the coupled system)
 and defines the choice of chiral symmetry operator (up to a sign).
The latter determines the total winding number. In the weak-coupling limit, it may range between the sum and differences of the winding numbers of individual chains.

Note that the other symmetries of the coupling, such as particle-hole and time-reversal symmetries are not relevant - as the winding number of a weakly coupled system is determined only by the chiral symmetry.  From this,  we conclude that the $\mathbb{Z}$ classes (BDI, AIII, CII) are topologically equivalent in one dimension as far as gapped systems are concerned.  The only difference between the edge states in these models is related to the symmetries that may or may not be there -- for example if the model has time-reversal symmetry, so will the edge states, but they will be equivalent to those of a model without the time-reversal symmetry in every other way.

In these classes, we have proved that the edge states are all localised on a single sublattice -- and if there are multiple edge states on a given edge, they all must be localised on the same sublattice.  This is in contrast to the class DIII, where while it has chiral symmetry, the other symmetries force the winding number to be zero.  In this case, the Kramer's pair of edge states (on e.g. the left edge) has one on the A sublattice and the other on the B sublattice.  In this case, one requires time-reversal and particle-hole symmetry in order for the edge states to remain pinned to zero energy (the middle of the gap).

It is worth observing in the case of DIII that if one looked at the edge-state Hamiltonian alone, one finds a single Kramer's pair which can not be gapped by any perturbation respecting time-reversal symmetry.  One might then come to the erroneous conclusion that time-reversal symmetry alone is all that is needed to protect such an edge state in 1D, which is clearly not true as the AII class in 1D has a trivial topological classification.  A further symmetry is required to pin these edge states to zero-energy, which is not be seen in the edge-state Hamiltonian alone.

\section*{Outlook }
We plan to extend this approach to three dimensions, where we expect to be able to construct lattice realisations of models in all of the universality classes.
We expect that for the gapped phases the chiral symmetry and sign ambiguity in winding number will play the same crucial role as they play in one dimension.

We also plan to study the effect of interactions on the topological properties of these 1D systems. The set of toy models constructed above will serve as 
a convenient framework from which interactions may be added. We will employ this to study realistic systems that can not be mapped onto 
standard spin-chain models studied in the literature \cite{Fidkowski2010, Fidkowski2011,PollmannVerresen2017}.

\section*{Acknowledgements}
D. G. is supported by ISF-China 3119/19 and ISF 1355/20.
P.M.  acknowledges support by the Israel Council for Higher Education Quantum Science and Technology Scholarship.

\appendix
\onecolumngrid
\section*{Appendix}
\section{Proof of Eq. (\ref{phase_detq})}\label{AppendixA}
In order to prove Eq. (\ref{phase_detq}) we consider an arbitrary $2N$ band one-dimensional chiral Hamiltonian brought to block-off diagonal form: 
\begin{align}
\label{H_chiral}
\hat{H}=\begin{pmatrix}
0 & \hat{\Delta} \\
\hat{\Delta}^{\dagger} & 0
\end{pmatrix}
\end{align}
Next we go into basis where $\hat{\Delta}$ is diagonal: 
\begin{equation}
\label{Delta_diagonal}
\hat{\Delta}=\begin{pmatrix}
 \epsilon_1 e^{i \theta_1}& 0 & 0 & \dots  \\
0 & \epsilon_2 e^{i \theta_2} & 0& \dots & \\
0& 0 & \epsilon_3 e^{i \theta_3} &  \dots& \\
\vdots & \vdots & \vdots & \ddots& \\
\end{pmatrix} 
\end{equation}
where $\epsilon_i>0$. The eigenstates of (\ref{H_chiral}) can be constructed as: 
\begin{align}
\label{H_eigenstates}
\Psi^{\pm}_j=\frac{e^{i\alpha}}{\sqrt{2}} \begin{pmatrix}
\chi_j \\ \pm e^{-i\theta_j} \chi_j, \\
 \end{pmatrix},\\
\end{align}
where $\chi_j$ is a unit $N$ component vector with $\chi^i_j= \delta_{ij}$ and $\alpha$ is an arbitrary phase. 
The eigenstates $\Psi^{\pm}_j$ correspond to eigenvalues $\pm\epsilon_j$. Next we construct the projector onto a band with energy $-\epsilon_a$ using (\ref{H_eigenstates}). For $a=1$ it has the following structure:
\begin{align}
\label{projector_1}
P_1=(\Psi^{-}_1)^{\dagger} \Psi^{-}_1=\frac{1}{2} \begin{pmatrix}
1 & 0 & \dots & -e^{i\theta_1} & 0 &\dots  \\
0 & 0 & \dots & 0 & 0 & \dots  \\
\vdots& \ddots &\vdots & \vdots& \vdots \\
-e^{-i\theta_1}  & 0 & \dots & 1& 0 & \dots\\
0 &0 &\dots & 0& 0 & \dots \\
\vdots &\vdots & \vdots & \vdots & \vdots & \vdots \\
\end{pmatrix}. 
\end{align} 
Thus for an arbitrary $a$ the non-zero elements are $P^{aa}_a=P^{a+N, a+N}_a=1$ and $P^{a, N+a}_a=(P^{N+a,a}_a)^*=e^{i\theta_a}$. Now we sum up over all the filled states (states with negative energy) and get: 
\begin{align}
\label{Projector_sum}
P=\sum_a (\Psi^{-}_a)^{\dagger} \Psi^{-}_a=\frac{1}{2} \begin{pmatrix}
\text{I}_N & -\bar{\Delta} \\
-\bar{\Delta}^* & \text{I}_N
\end{pmatrix},
\end{align}
where $\text{I}_N$ is an identity matrix and block $\bar{\Delta}$ is given by: 
\begin{equation}
\label{bar_delta}
\bar{\Delta}=\begin{pmatrix}
 e^{i \theta_1}& 0 & 0 & \dots  \\
0 & e^{i \theta_2} & 0& \dots & \\
\vdots& \vdots & \ddots &  \vdots& \\
0 & \dots & \dots & e^{i\theta_N}& \\
\end{pmatrix} 
\end{equation}
In order to construct the winding number, we follow \cite{Ryu2010} and consider the operator $Q$ expressed through the projector $P$ onto the filled bands: $Q(k)=1-P(k)$.  
This operator is chiral symmetric, i.e. $\{C, Q\}=0$, thus in the basis where the operator $C$ is block-diagonal i.e. $C= \hat{\lambda} \sigma_z$, where $\hat{\lambda}$ is some unitary matrix, the matrix $Q(k)$ has the block off-diagonal form: 
\begin{align}
\label{Qmatrix_off}
Q(k) = \begin{pmatrix}
0 & q(k) \\ 
q^{\dagger}(k) & 0 \\
\end{pmatrix}. 
\end{align}
In our case, the block $q(k)= \bar{\Delta}$ as follows from (\ref{Projector_sum}). 
The determinant of the block $q(k)=\bar{\Delta}$ is given by: 
\begin{align}
\label{det_q}
\text{det}[q(k)]=\exp\left[i\sum\limits_j \theta_j \right] \equiv e^{i \phi}, 
\end{align}
As $\tr [q^{-1} \partial_k q]=\partial_k \ln \text{det} [q(k)] = i \partial_k \phi $ we obtain the following expression for the winding number: 
\begin{equation}
\label{winding_phi}
\nu=\frac{i}{2\pi} \int_{\text{BZ}} dk \tr [q^{-1} \partial_k q]=- \frac{1}{2\pi} \int_{\text{BZ}} dk \partial_k \phi. 
\end{equation}
Now we express the phase $\phi$ through the determinant of $q$: 
\begin{equation}
\label{q_phase}
\phi =\arctan\left[\frac{\text{Im} \det[q(k)]}{\text{Re} \det[q(k)]}\right]
\end{equation}
 As according to (\ref{Delta_diagonal}) $\det [\Delta(k)]= \prod\limits_j \epsilon_j e^{i \phi} $, we can rewrite the expression (\ref{q_phase}) in terms of $\det[\Delta(k)]$ by dividing and multiplying the argument of arctan by $\prod\limits_j \epsilon_j$ and get: 
\begin{equation}
\label{Eq9}
\phi =\arctan\left[\frac{\text{Im} \det[\Delta(k)]}{\text{Re} \det[\Delta(k)]}\right]
\end{equation}
Rewriting this expression in a slightly more physically transparent way gives us Eq.~\eqref{phase_detq}.  
\section{Sign ambiguity of a winding number in odd dimensions} \label{AppendixB}
Here we prove that the winding number for the $N$-band chiral model in odd dimensions is defined up to a sign.
To do that, we consider an expression for a winding number in an odd-dimensional $d=2n+1$ space \cite{Ryu2010}: 
 \begin{align}
 \label{winding number_d}
 \nu=\int_{\text{BZ}^{d=2n+1}} \omega_{2n+1},
 \end{align}
 where the winding number density $\omega_{2n+1}$ is defined as: 
 \begin{align}
 \label{wd_density}
 \omega_{2n+1}=\frac{(-1)^nn!}{(2n+1)!} \left(\frac{i}{2\pi}\right)\epsilon^{\alpha_1\alpha_2...\alpha_d}\tr\left[q^{-1}\partial_{\alpha_1}q\cdot q^{-1}\partial_{\alpha_2}q...q^{-1}\partial_{\alpha_d}q\right] d^{2n+1}k,
 \end{align}
 where $\epsilon^{\alpha_1\alpha_2...\alpha_d}$ is a $d$ dimensional Levi-Civita symbol and $\partial_{\alpha_i} \equiv \partial_{k_i}$. 
We remind the reader, that the matrix $q$ is a block of another chiral symmetric hermitian matrix $Q$, constructed through a projector onto filled bands: 
\begin{align}
\label{Qmatrix_off_2}
Q(k) = \begin{pmatrix}
0 & q(k) \\ 
q^{\dagger}(k) & 0 \\
\end{pmatrix}. 
\end{align}
Here $k={k_1,k_2,...,k_d}$. This matrix is written is the basis $\Psi^{\text{T}}=\{\Psi_1,\Psi_2,...\Psi_{N/2},\Psi_{N/2+1},...\Psi_N\}$. Now, we can re-arrange the components of the spinor, and write it as $\Psi^{\text{T}}\rightarrow \Psi^{\text{T}}=\{\Psi_{N/2+1},...\Psi_N,\Psi_1,\Psi_2,...\Psi_{N/2}\}$. This corresponds to the unitary transformation of the form $U=\sigma_x \hat{I}$, where $\hat{I}$ is a $\frac{N}{2}\cross \frac{N}{2}$ unit matrix. In the new basis the matrix $Q$ is still off block-diagonal, however, under this transformation its block transforms as $q \rightarrow q^{\dagger}$. Matrix $q$ is unitary, therefore: 
\begin{align}
\label{qdq}
\partial_{\alpha_i}\left(q^\dagger q\right)=q^{\dagger} \partial_{\alpha_i} q+q \partial_{\alpha_i} q^{\dagger}=0.
\end{align}
This implies: 
\begin{align}
\tr\left[q\partial_{\alpha_1}q^{-1}\cdot q \partial_{\alpha_2}q^{-1}...q\partial_{\alpha_d}q^{-1}\right]=(-1)^{d}\tr\left[q^{-1}\partial_{\alpha_1}q\cdot q^{-1} \partial_{\alpha_2}q...q^{-1}\partial_{\alpha_d}q\right]. 
\end{align}
We apply this property to the winding number density and use that as $d$ is odd,i.e. $(-1)^{d}=-1$ and therefore we prove that the winding number density (\ref{wd_density}) and correspondingly the winding number (\ref{winding number_d}) change the sign under a unitary transformation of a basis. 

\section{General models}
\label{AppendixC}
\subsection{Construction of models in k-space}
The general Hamiltonian that describes coupled chains is: 
\begin{align}
\hat{H}=\begin{pmatrix}
\hat{h}_{\text{SSH}} & \hat{W} \\
\hat{W}^{\dagger} & \hat{h}^*_{\text{SSH}}
\end{pmatrix},
\end{align} 
In order to construct models that represent chiral symmetric topological classes,  we study how the coupling matrix $\hat{W}$ transforms under symmetry operations (\ref{time-reversal}),  (\ref{particle-hole}).  For simplicity we do that in $k-$ space. 
We obtain the following properties on the matrix $\hat{W}(k)$ by imposing symmetry constraints and taking into account that in Fourier space the operator $\mathcal{K}$ reverses the sign of momentum $k \rightarrow -k$:   
\begin{align*}
T_-: \hspace{0.5cm} \hat{W}(k)=-\hat{W}^{\text{T}}(-k) \nonumber \\
T_+: \hspace{0.8cm} \hat{W}(k)=\hat{W}^{\text{T}}(-k) \nonumber \\
P_-: \hspace{0.5cm} S_z \hat{W}(k)S_z=\hat{W}^{\text{T}}(-k) \nonumber \\
P_+: \hspace{0.45cm} S_z\hat{W}(k)S_z=-\hat{W}^{\text{T}}(-k). \nonumber 
\end{align*} 
By taking into account these conditions, we obtain the general form of the matrix $\hat{W}(k)$ for symmetry classes with chiral symmetry, see the Table \ref{V_classes}. 
\begin{table}[h]
\begin{tabular}{lclclc|c|}
 Class & $T^2$ & $P^2$ & $\hat{W}(k)$ \\
 \hline
BDI & 1 & 1 & $f_e(k) S_x+f_o(k)S_y$  \\
CII & -1 & -1  & $f_o(k) S_x+f_e(k)S_y$ \\
DIII & -1 &1 & $f_o(k) S_0+g_0(k)S_z$ \\
CI & 1 & -1  & $f_e(k) S_0+g_e(k)S_z$\\
\hline
\end{tabular}
\caption{Momentum space structure of the coupling matrix $\hat{W}$ (\ref{block_H}) in different classes.  Here $f_e(k), g_e(k)$ are arbitrary even functions of $k$ and $f_o(k), g_o(k)$ -- arbitrary odd functions of $k$. }
\label{V_classes}
\end{table}
The real-space structure of the matrix $\hat{W}$ depends on the choice of the even and odd functions $f_e(k), f_o(k), g_e(k), g_o(k)$.  If we focus on hopping terms up to nearest-neighbor,  the possible $k-$ dependence of the odd functions $f_o(k), g_o(k)$ is $\sin(k)$ and for the even functions $f_e(k), g_e(k)$ we can choose either constant (corresponds to on-site terms) or $\cos(k)$.    
\subsection{Classes BDI and CII}
The general Hamiltonian belonging to the class AIII with chiral symmetry $C_1=S_z\sigma_0$ reads: 

\begin{align}
\label{H_C1_general}
\hat{H}_{1} = \hat{H}_0+ \hat{V}_{\text{1}}, \nonumber \\ \hat{V}_{\text{1}}= \sum_n \hat{c}_{A,n}^{\dagger} \left [\vb{v} \cdot \vb*{\sigma} \right] \hat{c}_{B,n}+ \hat{c}_{B,n}^{\dagger} \left [\vb*{\omega}  \cdot \vb*{\sigma} \right] \hat{c}_{A,n+1} 
+ \text{h.c.}, \\
\vb{v} =\{v_x,v_y,0\},  \hspace{0.5cm} \vb*{\omega} =\{\omega_x,\omega_y,0\}, \nonumber
\end{align}
where $\vb*{\sigma}$ is the vector of three Pauli matrices acting in chain basis.  The case of real coupling amplitudes corresponds to the topological class BDI, while the case of imaginary couplings describes the class CII.  By setting $v_y=\omega_y=0$ and $v_x=\omega_x=a$ we obtain the minimal model (\ref{BDI_CII}) that we studied in the main text. 

\subsection{Classes DIII and CI}

Now consider the general Hamiltonian belonging to the class AIII with chiral symmetry $C_2=S_z\sigma_z$:
 \begin{align}
\label{H_C2_general}
\hat{H}_{2} = \hat{H}_0+ \hat{V}_{\text{2}}, \nonumber \\ \hat{V}_{\text{2}}=
\sum_n \hat{c}_{A,n}^{\dagger} \left [\vb*{\beta}_A \cdot \vb*{\sigma} \right] \hat{c}_{A,n+1}+ \hat{c}_{B,n}^{\dagger} \left [\vb*{\beta}_B  \cdot \vb*{\sigma} \right] \hat{c}_{B,n+1} +
  \sum_n \hat{c}_{A,n}^{\dagger} \left [\vb*{\delta}_A \cdot \vb*{\sigma} \right] \hat{c}_{A,n}+ \hat{c}_{B,n}^{\dagger} \left [\vb*{\delta}_B  \cdot \vb*{\sigma} \right] \hat{c}_{B,n} 
+ \text{h.c.}, \\
\vb*{\beta}_{A/B} =\{\beta_{A/B,x} , \beta_{A/B,y} ,0\},  \hspace{0.5cm} \vb*{\delta}_{A/B} =\{\delta_{A/B,x},\delta_{A/B,y},0\}.\nonumber
\end{align}
The case of imaginary amplitudes $\vb*{\beta}_{A/B}$ and $\vb*{\delta}_{A/B} = 0$ corresponds to the class DIII and the case of real amplitudes $\vb*{\beta}_{A/B}$ describes the model of trivial topological class CI. 

\section{Edge states}\label{appendix_edge}
In order to obtain the edge states solution, we use the Heisenberg picture.  In this picture,  the fermionic creation operator $\hat{c}^{\dagger}_{A/B,m, \sigma}$ obeys the time evolution determined by the commutator with the Hamiltonian of a model: 
\begin{align}
\label{Heisenberg}
-i\frac{d}{dt} \hat{c}^{\dagger}_{A/B,m, \sigma}= [\hat{H},  \hat{c}^{\dagger}_{A/B,m, \sigma}],
\end{align}
where the operators $\hat{c}^{\dagger}_{A/B,m, \sigma}$ are related to the components of the wavefunction $\Psi_{A/B,m,\sigma}$ as: 
\begin{align}
\hat{c}^{\dagger}_{A/B,m, \sigma} =\sum\limits_{\epsilon} \Psi_{A/B,m,\sigma} e^{i\epsilon t} \hat{c}^{\dagger}_{\epsilon,A/B,\sigma} \equiv \sum\limits_{\epsilon} \hat{c}^{\dagger}_{\epsilon,A/B,\sigma}(m)
\end{align}
If we substitute this ansatz to the Heisenberg equation (\ref{Heisenberg}),  we obtain the following stationary equations: 
\begin{align}
\label{Heisenberg_stationary}
\epsilon \hat{c}^{\dagger}_{\epsilon,A/B,\sigma}(m)= [\hat{H},  \hat{c}^{\dagger}_{\epsilon,A/B,\sigma}(m)]
\end{align}
From here one can obtain the corresponding wavefunction $\Psi_{A/B,m,\sigma}$. 
\subsection{Edge states in CII and BDI classes }
The Hamiltonian with the chiral symmetry $C_1$ is given by (\ref{H1}),(\ref{BDI_CII}): 
\begin{align}
\label{C1_edge}
H_{1} =w\sum_{n=1}^{N}c^{\dagger}_{An1}c_{Bn1} + v\sum_{n=1}^{N-1}c^{\dagger}_{Bn1}c_{A,n+1,1}  + w^*\sum_{n=1}^{N}c^{\dagger}_{An2}c_{Bn2} + v^*\sum_{n=1}^{N-1}c^{\dagger}_{Bn2}c_{A,n+1,2}+ \nonumber  \\
+ a\sum_{n}(c^{\dagger}_{An1}c_{Bn2} + c^{\dagger}_{Bn1}c_{A,n+1,2}+c^{\dagger}_{An2}c_{Bn1} + c^{\dagger}_{Bn2}c_{A,n+1,1})+ \text{h.c}.
\end{align}
If $a$ is real the Hamiltonian describes a model that belongs to BDI class and if $a$ is imaginary,  the model belongs to CII class.   We will focus on the topological phase with $\nu=2$, i.e. there are two edge states. The equation of motion for this model (\ref{Heisenberg_stationary}) yields the following equation for the wavefunction of the model: 
\begin{align}
\label{equations_C1}
\begin{cases}
 w^{*} \Psi_{Am1} +v  \Psi_{A,m+1,1}+ a\Psi_{A,m+1,2} + a^{*}\Psi_{Am2} = \epsilon \Psi_{Bm1}\\
w^{*} \Psi_{Am2}+ v\Psi_{A,m+1,2} + a\Psi_{A,m+1,1} +a^{*} \Psi_{Am1} = \epsilon \Psi_{Bm2}\\
 w \Psi_{Bm1} +v^{*} \Psi_{B,m-1,1} + a\Psi_{Bm2} + a^{*} \Psi_{B,m-1,2}= \epsilon  \Psi_{Am1} \\
 w\Psi_{Bm2} + v^{*} \Psi_{B,m-1,2} +a\Psi_{Bm1} + a^{*} \Psi_{B,m-1,1}=  \epsilon  \Psi_{Am2} 
\end{cases}
\end{align}
If we are looking for the midgap states ($\epsilon=0$) the equations for sublattices $A$ and $B$ decouple.  
As we discussed in the main text,  the two protected edge states should be localised on the same sublattices. Therefore,  if we consider a half-infinite system, we can look for solutions localised on the left edges.  Thus atoms A on the left edge decouple and $\Psi_{B,m,\sigma}=0$.  In this case we are dealing with the following system: 
\begin{align}
\label{edge_states_A}
v\Psi_{A,m+1,1} + a\Psi_{A,m+1,2} &= -( w^{*}\Psi_{Am1}+a^{*}\Psi_{Am2})\\
a\Psi_{A,m+1,1} +v^{*} \Psi_{A,m+1,2} &= -(a^{*} \Psi_{Am1}+ w\Psi_{Am2})
\end{align}
what we have produced here is a recurrence relation that gives us the wavefunction for the next cell based on its value on the current cell.
Express as matrices,

\begin{equation}
\label{Matrix_edge_states}
  \begin{pmatrix} v & a \\ a & v^{*} \end{pmatrix} \begin{pmatrix} \Psi_{A,m+1,1}\\ \Psi_{A,m+1,2} \end{pmatrix} = -\begin{pmatrix} w^{*} & a^{*}\\ a^{*} & w \end{pmatrix} \begin{pmatrix} \Psi_{Am1}\\ \Psi_{Am2} \end{pmatrix}
\end{equation}
By introducing the matrix C that represents the 2x2 matrix on the left side,  and the matrix D on the right side,  we obtain:
\begin{align}
  \underline{\psi}_{A,m+1}=-C^{-1}D\underline{\psi}_{Am} 
\end{align}
we can define the "transfer" matrix T,  by combining C and D matrices $T=C^{-1}D$:
\begin{align}
  \underline{\psi}_{A,m+1}=-T\underline{\psi}_{Am},  \nonumber \\
  T=\frac{1}{|v|^{2}-a^{2}}\begin{pmatrix} w^{*}v^{*}-|a|^{2} & v^{*}a^{*}-a w\\ va^{*}-w^{*}a & wv-|a|^{2} \end{pmatrix}
\end{align}
In order to construct the exponentially decaying states we find the eigenvalues of the transfer matrix: 
\begin{align}
\lambda_{1,2}=\frac{1}{2}(-(2|a|^{2}-w^{*}v^{*}-wv) \pm i\sqrt{|\Omega|}),  \nonumber \\
\Omega=8|a|^2 \text{Re}[wv] -2\text{Re}[(wv)^2]+2|w|^2|v|^2-4(|v|^2(a^*)^2+|w|^2a^2)
\end{align}
We can also define the logarithm of those eigenvalues $\delta =\log (\lambda)$.  Real part of $\delta$ describes the decaying length of the edge states and the imaginary part describes the oscillating part of the wavefunction.  Note that at "high symmetry points", where the parameter $a$ is real or imaginary,  the eigenvalues are related by complex conjugation: $\lambda_1=\lambda_2^*$. 
The eigenvectors $\underline{u}_{1,2}$ of the transfer matrix are given by:
\begin{align}
\underline{u}_{1,2}=\left[1 \hspace{0.5cm} \frac{wv-w^*v^*\pm i\sqrt{|\Omega|}}{2(v^*a^*-aw)}\right].
\end{align} 
The most generic solution of the equation (\ref{Matrix_edge_states}) can be written as: 

\begin{equation}
  \underline{\psi}_{A,m}=\sum_{i}\beta_{i}(-\lambda_i)^{m}\underline{u_{i}},
\end{equation}
where $\beta_{1/2} \in \mathbb{C}$.
Expanding,

\begin{align}
\underline{\psi}_{A,m}=\begin{pmatrix} \Psi_{Am1}\\\Psi_{Am2}\end{pmatrix} = \beta_{1}(-\lambda_{1})^{m} \underline{u}_1+ \beta_{2}(-\lambda_{2})^{m} \underline{u}_2
\end{align}
Since both eigenstates are degenerate,  we can take any linear combination however the choice of a symmetric and antisymmetric combinations is useful to demonstrate some properties of the edge states.
In order to do that consider $\beta_{1}=\pm\beta_{2}$, and have the following definitions,
\begin{equation}
  \begin{aligned}
    \underline{\psi}_{+}&\Rightarrow\beta_{1}=\beta_{2}\\
    \underline{\psi}_{-}&\Rightarrow\beta_{1}=-\beta_{2}
  \end{aligned}
\end{equation}
Refine the equation to a simpler notation where the minus has been incorporated into the eigenvalues ($\lambda_{i}$):
\begin{equation}
    \underline{\psi}_{\pm} = (\lambda_{1})^{n} \underline{u}_{1} \pm (\lambda_{2})^{n} \underline{u}_{2}
\end{equation}
Note that the states are normalizable if they decay into the bulk,  i.e $|\lambda_{1,2}|<1$.

The symmetry properties of the model (\ref{C1_edge}) must be reflected in the properties of the edge state wavefunctions. We can demonstrate analytically the properties of these states by studying how they transform under the action of time-reversal symmetry. 
Our conjecture is that for the BDI class with the time-reversal symmetry $T_+^2=+1$,  application of the TRS will transform the state back to itself.  For the CII class, to comply with Kramers theorem, the state will transform to its counterpart:
\begin{equation}
\label{edge_state_conjecture}
  \begin{matrix} \text{BDI} & T_{+}\underline{\psi}_{\pm} \propto \underline{\psi}_{\pm} & T_{+}=S_{0}\sigma_{x}K\\
    \text{CII} & T_{-}\underline{\psi}_{\pm} \propto \underline{\psi}_{\mp} & T_{-}=iS_0\sigma_{y}K
    \end{matrix}
\end{equation}
Let us demonstrate those properties explicitly.  To do that we study how the time-reversal symmetry acts on the eigenvectors $\underline{u}_{1,2}$: 
\begin{equation}
    \begin{aligned}
        T_{+}\underline{u}_{1}&=S_{0}\sigma_{x}K\begin{pmatrix}1 \\ \frac{w^{*}v^{*}-wv + i\sqrt{|\Omega|}}{2(v^{*}a^{*}-wa)} \end{pmatrix}\\
        &=\begin{pmatrix}\frac{wv-w^{*}v^{*} - i\sqrt{|\Omega|}}{2(va-w^{*}a^{*})}\\ 1 \end{pmatrix}=\underline{u}_{2}
    \end{aligned}
\end{equation}

It is easy to check then that $T_{+}\underline{\psi}_{-}=-\underline{\psi}_{-}$. Thusly $T_{+}\underline{\psi}_{\pm}=\pm\underline{\psi}_{\pm}$ which is consistent with our conjecture (\ref{edge_state_conjecture}).

By acting $T_-$ on the eigenstates $\underline{u}_1$ and $\underline{u}_2$ we get: 
\begin{equation}
    \begin{aligned}
        T_{-}\underline{u}_{1}&=iS_{0}\sigma_{y}K\begin{pmatrix}1 \\ \frac{w^{*}v^{*}-wv + i\sqrt{|\Omega|}}{2(v^{*}a^{*}-wa)} \end{pmatrix}\\
        &=\begin{pmatrix}\frac{wv-w^{*}v^{*} - i\sqrt{|\Omega|}}{2(va-w^{*}a^{*})}\\ -1 \end{pmatrix}=-\underline{u}_{2}
    \end{aligned}
\end{equation}
Thus our conjecture (\ref{edge_state_conjecture}) also holds for the time reversal symmetry of CII class since operating on one of the eigenvectors gives a minus sign needed to transform $\underline{\psi}_{+}$ to $\underline{\psi}_{-}$. This follows from the fact that operating twice on the state should return the negative of the original state.  Therefore the characteristic feature of CII class is that the edge states can be chosen to form a Kramers doublet.  If the time-reversal symmetry is broken,  the eigenvectors are not related to each other by any symmetry transformation.  This is illustrated in the Fig.  \ref{figure_edge_states} of the main text. 
\subsection{Edge states in DIII class and their protection}
Here we derive the edge states of the model that has the chiral symmetry $C_2$: 
\begin{align}
\label{edge_H2}
H_{\text{2}} =w\sum_{n=1}^{N}c^{\dagger}_{An1}c_{Bn1} + v\sum_{n=1}^{N-1}c^{\dagger}_{Bn1}c_{A,n+1,1}  + w^*\sum_{n=1}^{N}c^{\dagger}_{An2}c_{Bn2} + v^*\sum_{n=1}^{N-1}c^{\dagger}_{Bn2}c_{A,n+1,2}+ \nonumber  \\
+b \sum_{n=1}^{N-1} \left( c^\dagger_{B,1,n} c_{B,2,n+1} + c^\dagger_{B,2,n} c_{B,1,n+1} 
 + c^\dagger_{A,1,n} c_{A,2,n+1} 
 + c^\dagger_{A,2,n} c_{A,1,n+1} \right) + h.c.,
\end{align}
where $b=|b|e^{i\phi}$.  If $\phi=\pi/2$,  the Hamiltonian (\ref{edge_H2}) has time-reversal symmetry and belongs to the class DIII.  For any other $\phi \neq \pi/2$ the time-reversal symmetry is broken and the model belongs to the trivial class CI. 
Equations of motion for this model are: 
\begin{align}
\label{equations_CI}
\begin{cases}
w \Psi_{Bm1}+v^* \Psi_{B,m-1,1}+b\Psi_{A,m+1,2}+ b^*\Psi_{A,m-1,2} =\epsilon\Psi_{Am1} \\
 w^* \Psi_{Am1}+v \Psi_{A,m+1,1}+b \Psi_{B,m+1,2}+b^* \Psi_{B,m-1,2} =\epsilon\Psi_{Bm1} \\
 w^* \Psi_{Bm2}+v \Psi_{B,m-1,2}+b \Psi_{A,m+1,1}+b^*\Psi_{A,m-1,1} =\epsilon\Psi_{Am2}\\
 w \Psi_{Am2}+v^* \Psi_{A,m+1,2}+b\Psi_{B,m+1,1}+b^*\Psi_{B,m-1,1} =\epsilon\Psi_{Bm2}
\end{cases}
\end{align}
In this case the equations for sublattices $A$ and $B$ do not decouple if we focus on zero energies $\epsilon=0$.  However they decouple if we define new sublattices $A'$ and $B'$ according to the operator $C_2$ as $ A  \rightarrow  A',  B  \rightarrow  B'$ on the first chain and $ A  \rightarrow  B',  B  \rightarrow  A'$ on the second. 
 In order to solve the equations we take the following ansatz: 
\begin{align}
\label{ansatz_CI}
\underline{\psi}_m= \underline{\psi}_0 e^{\delta m}, \nonumber \\
\underline{\psi}_m = \begin{pmatrix}
\Psi_{A'm1}\\
\Psi_{A'm2}\\
\Psi_{B'm1} \\
\Psi_{B'm2} 
\end{pmatrix}, \underline{\psi}_0 = \begin{pmatrix}
c_{A',1}\\
c_{A',2} \\
c_{B',1} \\
c_{B',2} 
\end{pmatrix},
\end{align}
where the coefficients $c_{i}$ are complex numbers.  Note,  that for convenience we use here the parameter $\delta$ and not the exponent of it $\lambda=e^{\delta}$ as in the previous subsection.  As always,  real part of $\delta$ corresponds to the decaying length of the edge states (if $\delta<0$) and imaginary part describes the oscillating part.  We substitute (\ref{ansatz_CI}) to the equations (\ref{equations_CI}) and obtain two independent sets of equations: 
\begin{align}
(\romannumeral 1)
\begin{cases}
c_{B',1}(w+v^*e^{-\delta})+2c_{B',2}\cosh(\delta+i\phi)|b| =0 \\
c_{B',2}(w+v^*e^{\delta})+2c_{B',1}\cosh(\delta +i\phi)|b| =0
\end{cases}
(\romannumeral 2) 
\begin{cases}
c_{A',1}(w^*+v e^{\delta})+2c_{A',2}\cosh(\delta+i\phi) |b| =0 \\
c_{A',2}(w+v^*e^{-\delta})+2c_{A',1}\cosh(\delta +i\phi)|b| =0
\end{cases}.
\end{align}
We can write them in a compact matrix form, if we introduce two vectors $\underline{c}_1= (c_{B',1},  c_{B',2})$ and $\underline{c}_2= (c_{A',1},  c_{A',2})$.  With these notations the equations can be written as: 
\begin{align}
(\romannumeral 1):\hspace{0.5cm} \text{M}_1 \underline{c}_1= 0 \nonumber \\
(\romannumeral 2) :\hspace{0.5cm}  \text{M}_2 \underline{c}_2= 0,
\end{align}
where the matrices $\text{M}_1$ and $\text{M}_2$ are given by:
\begin{align}
\text{M}_1=\begin{pmatrix}
w+v^*e^{\delta} & 2|b|\cosh(\delta+i\phi) \\
2|b| \cosh(\delta+i\phi)& w+v^*e^{-\delta} 
\end{pmatrix} \\
\text{M}_2=\begin{pmatrix}
w^*+ve^{\delta} & 2|b|\cosh(\delta+i\phi) \\
2|b| \cosh(\delta+i\phi)& w^*+ve^{-\delta} 
\end{pmatrix}.
\end{align}
From the condition $\det \text{M}_{1,2} =0$ we can obtain $\delta$.  Let us write those two equations explicitly: 
\begin{align}
\label{equations_delta}
\det \text{M}_{1}= 0 \rightarrow w^2+(v^*)^2+2wv^*\cosh(\delta)-4|b|^2\cosh^2(\delta+i\phi)=0 \\
\det \text{M}_{2}= 0 \rightarrow (w^*)^2+(v)^2+2w^*v\cosh(\delta)-4|b|^2\cosh^2(\delta+i\phi)=0.
\end{align}
Our goal is to demonstrate that in the absence of time-reversal symmetry there are no zero-energy edge state solutions.  In order to do that we assume that the time-reversal symmetry is weakly broken,  so we can represent $\phi=\pi/2+ \alpha$, where $\alpha \ll 1$,  and we do perturbation theory in $\alpha$: 
\begin{align}
\label{pert_solution}
\delta = \delta_0+ \delta_{\alpha} + o(\alpha^2),  |\delta_{\alpha}| \ll 1
\end{align}
By substituting this ansatz to the equations (\ref{equations_delta}) for $\delta$,  we obtain the following expression for the correction $\delta_{\alpha}$: 
\begin{align}
\label{delta_alpha_correction}
\delta_{\alpha}=\frac{-4i|b|^2 x_0}{v^{*}w +4|b|^2x_0 } \alpha
\end{align}
Here we write the correction to the roots of $\det\text{M}_{1} =0$. The correction to $\det\text{M}_{2}=0$ can be obtain by replacing $v^*w \rightarrow v w^{*}$.   By $x_0 =\cosh[\delta_0]$ we denote the solutions for $\phi =\pi/2$, i.e. for the time-reversal symmetric case.  In the limit $\phi =\pi/2$ the equations (\ref{equations_delta}) become simple quadratic equations with the following solutions: 
\begin{align}
\label{solutions_tr_symmetric}
x_{0,1(2)}= \frac{-2vw \pm \sqrt{4w^2v^2 -16 |b|^2(-4|b|^2+v^2+w^2)}}{8|b|^2}, \hspace{0.2cm} x_{0,1(2)}=\cosh[\delta_{0,1(2)}]
\end{align}
As it is hard to work with a generic expression,  we can focus on a simple limit, where we know that the edge states exist.  This corresponds to the limit $|w|<|v|$ and $|b| < |v|-|w|$.  One can express explicitly $\delta_{0,(1)2}$ and obtain the following simple expressions: 
\begin{align}
\label{delta_tr}
\delta_{0, 1(2)} = \log [\pm i\frac{b}{v^*} ] \pm \frac{iw }{2b}
\end{align}
Similarly,  one can consider the equations for the sublattice $A'$ and corresponding $\delta_{0, 3(4)}= (\delta_{0,1(2)})^*$ as expected in time-reversal symmetric case.  
By using (\ref{delta_alpha_correction}) we construct the solution when the time-reversal symmetry is weakly broken.  In this case the solutions $\delta$ are not related by time-reversal symmetry anymore, so they do not form Kramers pairs.  The general solution of (\ref{equations_CI}) has the following form: 
\begin{align}
\label{solutions_CI}
\underline{\psi}_{m}= \beta_1 \underline{\psi}_{0,1} e^{\delta_1 m}+ \beta_2 \underline{\psi}_{0,2} e^{\delta_2 m} +  \beta_3 \underline{\psi}_{0,3} e^{\delta^*_1 m}+ \beta_2 \underline{\psi}_{0,4} e^{\delta^*_2 m}
\end{align}
Here the eigenvectors $\underline{\psi}_{0,i}$ (see the definition (\ref{ansatz_CI})) have the following structure: 
\begin{align}
\label{solutions_CI_spinors}
\underline{\psi}_{0,(1,2)}= \begin{pmatrix}
0\\
0\\
X_B(\delta_{1,2}) \\
1
\end{pmatrix}, \underline{\psi}_{0,(3,4)}= \begin{pmatrix}
X_A(\delta_{3,4})\\
1\\
0 \\
0
\end{pmatrix},
\end{align}
where $X_{A,B}(\delta)$ are given by: 
\begin{align}
X_A(\delta) = -\frac{2|b|\cosh(\delta + i\phi)}{w+v^*e^{\delta}} \nonumber \\
X_B(\delta)= -\frac{2|b|\cosh(\delta + i\phi)}{w^*+ve^{\delta}} 
\end{align}
The coefficients $\beta_i$ need to be chosen according to the boundary conditions at $m=0$. 
 Similarly to the case of BDI/CII classes,  at the boundary the wavefunction must vanish $\underline{\psi}_{m=0}=0$.  This can be satisfied if $\beta_1=-\beta_2$ and $\beta_3=-\beta_4$,  as follows directly from (\ref{solutions_CI}) and (\ref{solutions_CI_spinors}). Therefore,  $X_B(\delta_1)=X_B(\delta_2)$ or $X_A(\delta_3)=X_A(\delta_4)$.  One can check that for the time-reversal symmetric case those conditions are satisfied,  and when the symmetry is broken $\phi = \pi/2+\alpha$,  the difference between the two parts of the equality is non-zero and is given by (in the limit we are focused on): 
\begin{align}
\label{X_difference}
X_B(\delta_1)-X_B(\delta_2) = -\frac{2 i \alpha b w }{(v^*)^2}, \nonumber \\
X_A(\delta_3)-X_A(\delta_4) = -\frac{2 i \alpha b w^* }{v^2}. 
\end{align}
Therefore the boundary conditions for the edge states cannot be satisfied if the time-reversal symmetry is broken. 
Moreover,  one can demonstrate that in time-reversal symmetric case,  the edge states form Kramers pairs.   
We notice that $T_-\underline{\psi}_{0,1}=\underline{\psi}_{0,3}$ and $T_-\underline{\psi}_{0,3}=-\underline{\psi}_{0,1}$.  Similarly,  $T_-\underline{\psi}_{0,2}=\underline{\psi}_{0,4}$ and $T_-\underline{\psi}_{0,4}=-\underline{\psi}_{0,2}$. Therefore,  the following pair of states forms a Kramers pair (up to a normalization constant): 
\begin{align}
\label{Kramers_DIII}
\underline{\psi}_{\pm}=  (\underline{\psi}_{0,1} e^{\delta_1 m} - \underline{\psi}_{0,2} e^{\delta_2 m}) \pm (\underline{\psi}_{0,3} e^{\delta^*_1 m} - \underline{\psi}_{0,4} e^{\delta^*_2 m}). 
\end{align}

\bibliography{draft}
\end{document}